\documentclass{elsart}

\usepackage{amsmath,amssymb}
\usepackage{times,txfonts,bm}
\usepackage{graphicx}

\usepackage{color}
\usepackage{rotating}

\definecolor{FGViolet}{rgb}{0.61,0.32,0.61}
\definecolor{FGBlue}{rgb}{0,0,0.8}
\definecolor{FGGreen}{rgb}{0.2,0.7,0.2}
\definecolor{FGRed}{rgb}{0.8,0,0}
\definecolor{FGLightGray}{rgb}{0.85,0.85,0.85}
\definecolor{FGGray}{rgb}{0.6,0.6,0.6}

\newcommand{\linemediumsolid}[1][black]{\unitlength1ex
  ({\color{#1}\begin{picture}(6,1)
  \linethickness{0.4mm}
  \put(0,0.5){\line(1,0){6.0}}
  \end{picture}})\nolinebreak
}

\newcommand{\linemediumdashed}[1][black]{\unitlength1ex
  ({\color{#1}\begin{picture}(6,1)
  \linethickness{0.4mm}
  \put(0,0.5){\line(1,0){1.5}}
  \put(2.2,0.5){\line(1,0){1.5}}
  \put(4.4,0.5){\line(1,0){1.5}}
  \end{picture}})\nolinebreak
}

\newcommand{\linemediumdotted}[1][black]{\unitlength1ex
  ({\color{#1}\begin{picture}(6,1)
  \linethickness{0.4mm}
  \put(0,0.5){\line(1,0){0.4}}
  \put(0.9,0.5){\line(1,0){0.4}}
  \put(1.8,0.5){\line(1,0){0.4}}
  \put(2.7,0.5){\line(1,0){0.4}}
  \put(3.6,0.5){\line(1,0){0.4}}
  \put(4.5,0.5){\line(1,0){0.4}}
  \put(5.4,0.5){\line(1,0){0.4}}
  \end{picture}})\nolinebreak
}

\newcommand{\linemediumdashdot}[1][black]{\unitlength1ex 
  ({\color{#1}\begin{picture}(6,1)
  \linethickness{0.4mm}
  \put(0,0.5){\line(1,0){0.4}}
  \put(0.9,0.5){\line(1,0){1.5}}
  \put(2.9,0.5){\line(1,0){0.4}}
  \put(3.8,0.5){\line(1,0){1.5}}
  \put(5.8,0.5){\line(1,0){0.4}}
  \end{picture}})\nolinebreak
}

\newcommand{\eq}[1]{\begin{equation}#1\end{equation}}
\newcommand{\eqmulti}[1]{\begin{equation}\begin{split}#1\end{split}\end{equation}}

\newcommand{\ket}[1]{\ensuremath{\,|{#1}\rangle}}
\newcommand{\braket}[2]{\ensuremath{\langle{#1}|{#2}\rangle}}

\newcommand{\matrixe}[3]{\ensuremath{\langle{#1}|\,{#2}\,|{#3}\rangle}}

\newcommand{\comm}[2]{\ensuremath{[{#1},{#2}]}}
\newcommand{\hermi}[1]{\ensuremath{\{{#1}\}_{\text{H}}}}

\newcommand{\op}[1]{\ensuremath{\mathrm{#1}}}
\newcommand{\adj}[1]{\ensuremath{{{#1}}^{\dag}}}
\newcommand{\corr}[1]{\ensuremath{\hat{#1}}}

\newcommand{\ii}{\ensuremath{\mathrm{i}}}
\newcommand{\dd}{\ensuremath{\mathrm{d}}}

\renewcommand{\vec}[1]{\ensuremath{\bm{#1}}}

\newcommand{\cO}{\ensuremath{\op{c}}}
\newcommand{\ccO}{\ensuremath{\adj{\op{c}}}}
\newcommand{\gO}{\ensuremath{\op{g}}}

\newcommand{\qO}{\ensuremath{\op{q}}}
\newcommand{\rO}{\ensuremath{\op{r}}}
\newcommand{\tO}{\ensuremath{\op{t}}}
\newcommand{\vO}{\ensuremath{\op{v}}}

\newcommand{\CO}{\ensuremath{\op{C}}}
\newcommand{\CCO}{\ensuremath{\adj{\op{C}}}}

\newcommand{\HO}{\ensuremath{\op{H}}}
\newcommand{\JO}{\ensuremath{\op{J}}}
\newcommand{\OO}{\ensuremath{\op{O}}}
\newcommand{\PO}{\ensuremath{\op{P}}}
\newcommand{\RO}{\ensuremath{\op{R}}}
\newcommand{\SO}{\ensuremath{\op{S}}}
\newcommand{\TO}{\ensuremath{\op{T}}}
\newcommand{\VO}{\ensuremath{\op{V}}}

\newcommand{\PiO}{\ensuremath{\op{\Pi}}}
\newcommand{\PsiO}{\ensuremath{\op{\Psi}}}

\newcommand{\bV}{\ensuremath{\vec{b}}}

\newcommand{\rV}{\ensuremath{\vec{r}}}
\newcommand{\xV}{\ensuremath{\vec{x}}}

\newcommand{\sigmaV}{\ensuremath{\vec{\sigma}}}
\newcommand{\piV}{\ensuremath{\vec{\pi}}}
\newcommand{\rhoV}{\ensuremath{\vec{\rho}}}
\newcommand{\omegaV}{\ensuremath{\vec{\omega}}}

\newcommand{\AC}{\ensuremath{\mathcal{A}}}

\newcommand{\lOV}{\ensuremath{\vec{\op{l}}}}
\newcommand{\pOV}{\ensuremath{\vec{\op{p}}}}
\newcommand{\qOV}{\ensuremath{\vec{\op{q}}}}
\newcommand{\rOV}{\ensuremath{\vec{\op{r}}}}
\newcommand{\sOV}{\ensuremath{\vec{\op{s}}}}
\newcommand{\xOV}{\ensuremath{\vec{\op{x}}}}

\newcommand{\sigmaOV}{\ensuremath{\vec{\op{\sigma}}}}

\newcommand{\tensorO}{\ensuremath{\op{s}_{12}}}
\newcommand{\tensorLLO}{\ensuremath{\op{s}_{12}(\lOV,\lOV)}}
\newcommand{\tensorQQO}{\ensuremath{\op{s}_{12}(\qOV_{\Omega},\qOV_{\Omega})}}
\newcommand{\tensorRQO}{\ensuremath{\op{s}_{12}(\rOV,\qOV_{\Omega})}}
\newcommand{\tensorbarQQO}{\ensuremath{\bar{\op{s}}_{12}(\qOV_{\Omega},\qOV_{\Omega})}}
\newcommand{\spinorbitO}{\ensuremath{(\vec{\op{l}}\cdot\vec{\op{s}})}}
\newcommand{\orbitsqrO}{\ensuremath{\vec{\op{l}}\,{}^2}}

\newcommand{\Rm}{\ensuremath{R_-}}
\newcommand{\DRm}{\ensuremath{R'_-}}

\newcommand{\Rp}{\ensuremath{R_+}}
\newcommand{\DRp}{\ensuremath{R'_+}}
\newcommand{\DDRp}{\ensuremath{R''_+}}
\newcommand{\DDDRp}{\ensuremath{R'''_+}}

\newcommand{\Rpm}{\ensuremath{R_{\pm}}}
\newcommand{\Rmp}{\ensuremath{R_{\mp}}}

\newcommand{\UCOM}{\ensuremath{\textrm{UCOM}}}
\newcommand{\cm}{\ensuremath{\textrm{cm}}}
\newcommand{\elem}[2]{\ensuremath{{}^{#2}\text{#1}}}

\begin{document}

\begin{frontmatter}

\title{Nuclear Structure based on Correlated Realistic 
Nucleon-Nucleon Potentials} 

\author[TUD]{R. Roth},
\author[GSI]{T. Neff}, 
\author[TUD]{H. Hergert}, and 
\author[GSI]{H. Feldmeier}

\address[TUD]{Institut f\"ur Kernphysik, Technische Universit\"at
Darmstadt, 64289 Darmstadt, Germany}
\address[GSI]{Gesellschaft f\"ur Schwerionenforschung, Planckstr. 1, 64291
Darmstadt, Germany}

\begin{abstract} 

We present a novel scheme for nuclear structure calculations based on
realistic nucleon-nucleon potentials. The essential ingredient is the
explicit treatment of the dominant inter\-action-induced correlations by
means of the Unitary Correlation Operator Method (UCOM). Short-range
central and tensor correlations are imprinted into simple, uncorrelated
many-body states through a state-independent unitary transformation.
Applying the unitary transformation to the realistic Hamiltonian leads to
a correlated, low-momentum interaction, well suited for all
kinds of many-body models, e.g., Hartree-Fock or shell-model. We employ
the correlated interaction, supplemented by a phenomenological correction
to account for genuine three-body forces, in the framework of 
variational calculations with antisymmetrised Gaussian trial states
(Fermionic Molecular Dynamics). Ground state properties of nuclei up to
mass numbers $A\lesssim60$ are discussed. Binding energies, charge radii,
and charge distributions are in good agreement with experimental data. We
perform angular momentum projections of the intrinsically deformed 
variational states to extract rotational spectra. 

\end{abstract}


\end{frontmatter}

\clearpage
\section{Introduction}
\label{sec:intro}

The advent of realistic nucleon-nucleon (NN) potentials has created a
supreme challenge and opportunity for nuclear structure theory: the
\emph{ab initio} description of nuclei. Several families of  realistic or
modern NN potentials have been developed over the past three decades ---
among others the Argonne, Bonn, and Nijmegen potentials
\cite{WiSt95,Mach89,Mach01,StKl93}. The latest members of these families
reproduce the NN-scattering data and the deuteron properties with high
precision. All realistic potentials exhibit a quite complicated operator
structure with substantial contributions from non-central terms, most
notably the tensor and spin-orbit interaction. Besides these, explicit
momentum dependent terms or quadratic orbital angular momentum and
spin-orbit operators enter. The most recent potentials also employ
charge-symmetry and charge-independence  breaking terms to further improve
the agreement with the experimental phase-shifts.

Given a realistic NN-potential we are faced with the difficult task of
solving the nuclear many-body problem. Ideally, one would like to treat
the (non-relativistic) quantum many-body problem \emph{ab initio}, i.e.,
without introducing further approximations. So far, \emph{ab initio}
solutions utilising realistic potentials are restricted to light nuclei
with $A\lesssim12$. Extensive studies on ground state properties and
excitation spectra in this mass range have been performed using, e.g.,
Green's Function  Monte Carlo methods \cite{PiVa02,PuPa97} and the no-core
shell model \cite{CaNa02,NaVa00b}. 

The results of these \emph{ab initio} calculations clearly show that in
addition to the realistic NN-potential a three-nucleon force is inevitable
to reproduce the experimental data on light nuclei. The Argonne group has
constructed a series of phenomenological three-nucleon potentials
\cite{PiPa01} to fit the binding energies and spectra of light nuclei to
experiment. In this way the results for ground states and low-lying
excitations are in good agreement with the experimental findings in the
whole accessible mass range. Recent developments in chiral effective field
theories \cite{EnMa03} provide a framework for a consistent derivation of
two-, three- and multi-nucleon forces from more fundamental grounds. 

Our aim is to describe the structure of larger nuclei on the basis of
realistic nucleon-nucleon interactions while staying as close as possible
to an \emph{ab initio} treatment of the many-body problem. The step
towards larger particle numbers requires a truncation of the full
many-body Hilbert space to a simplified subspace of tractable size. In the
simplest approach, the many-body state is described by a single Slater
determinant, as, e.g., in the Hartree-Fock approximation. More elaborate
approximations, the multi-configuration shell-model for example, allow for
many-body states which are represented as a superposition of several
Slater determinants. When combining realistic nucleon-nucleon interactions
with simple many-body states composed of single or few Slater determinants
a fundamental problem arises. Those states do not allow for an adequate
description of the strong short-range correlations induced by the
realistic NN-potential \cite{FeNe98,NeFe03}.

\begin{figure}
  \begin{center}
  \includegraphics[width=0.7\textwidth]{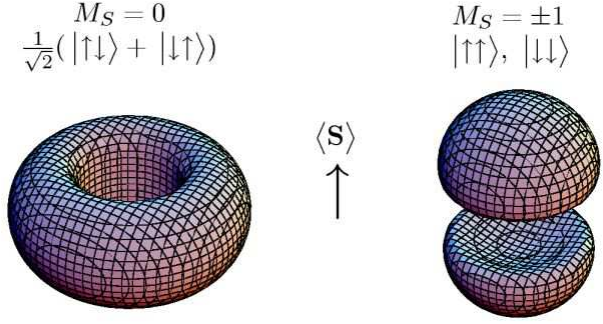}
  \end{center}
  \caption{Spin-projected two-body density $\rho^{(2)}_{1,M_S}(\rV)$ of the
  deuteron calculated with the AV18 potential. Shown is the iso-density
  surface for $0.005\;\textrm{fm}^{-3}$.}
  \label{fig:deuterondens}
\end{figure}

Already in the deuteron, structure and origin of these correlations are
apparent. We consider the spin-projected two-body density matrix
$\rho^{(2)}_{S=1,M_S}(\rV)$ as function of the relative coordinate
$\rV=\xV_1-\xV_2$ of the two particles, resulting from an exact calculation
with the Argonne V18 potential (AV18) \cite{WiSt95,FoPa96,NeFe03}. Figure
\ref{fig:deuterondens} depicts iso-density cuts of the two-body density
for $M_S=0$ and $M_S=\pm1$, respectively, which corresponds to parallel and
antiparallel alignment of the two nucleon spins. Two dominant structural
features appear: (\emph{i}) At small particle distances $|\rV|$ the
two-body density is fully suppressed, i.e., the probability of finding two
nucleons closer than $r \sim 0.5\, \text{fm}$ is practically zero.
(\emph{ii}) The angular structure of $\rho^{(2)}_{S=1,M_S}(\rV)$ depends
strongly on the spin orientation. For parallel spins the probability
density is concentrated along the quantisation axis (``dumbbell''), for
antiparallel spins it is constricted around the plane perpendicular to
the spin direction (``doughnut'').

These structures are the manifestation of two types of interaction-induced
correlations which govern the nuclear many-body problem: (\emph{i})
short-range central correlations and (\emph{ii}) tensor correlations.

\begin{figure}
  \begin{center} 
    \includegraphics[width=1.0\textwidth]{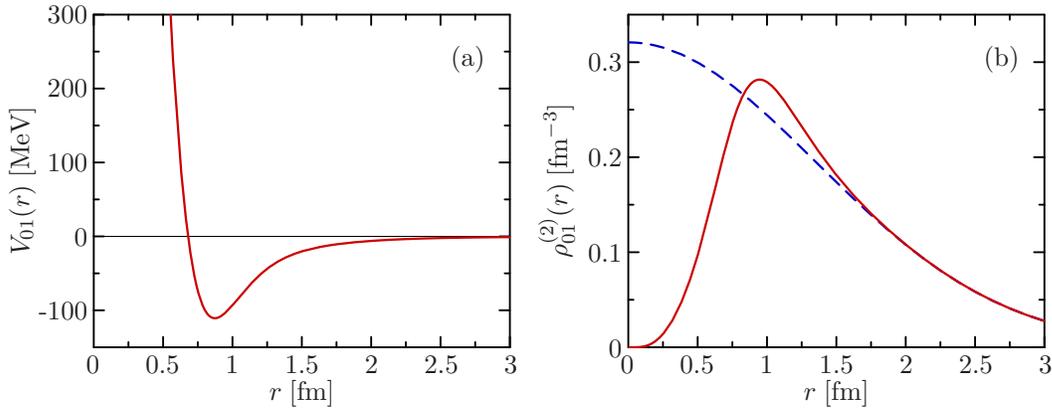}
  \end{center} 
  \caption{(a) Central part of the AV18 potential in the $(S,T)=(0,1)$
  channel. (b) Two-body density distribution $\rho^{(2)}_{S=0,T=1}(\rV)$
  for \elem{He}{4} resulting from an ansatz wave function using a Slater
  determinant of Gaussian single-particle states \linemediumdashed[blue]\
  compared to a realistic two-body density including the
  interaction-induced central correlations \linemediumsolid[red].}
  \label{fig:centralcorrillu}
\end{figure}

The central correlations are induced by the strong repulsive core of the
central part of the interaction. Figure \ref{fig:centralcorrillu} shows
the radial dependence of the central part of the AV18 potential in a
specific spin-isospin channel. Due to the short-range repulsion the
two-body density, shown for \elem{He}{4}, is completely suppressed
within the region of the repulsive core. A typical ansatz for the
many-body wave function using a Slater determinant of Gaussian (or any
other) single-particle states is not capable of describing this
correlation hole. The dashed line in Fig. \ref{fig:centralcorrillu}(b) shows
that the Slater determinant ansatz leads to a significant probability of
finding the nucleons within the core. Even with a superposition of a
moderate number of Slater determinants one is not able to describe the
strong short-range central correlations caused by realistic potentials.

\begin{figure}
  \begin{center} 
    \includegraphics[width=0.7\textwidth]{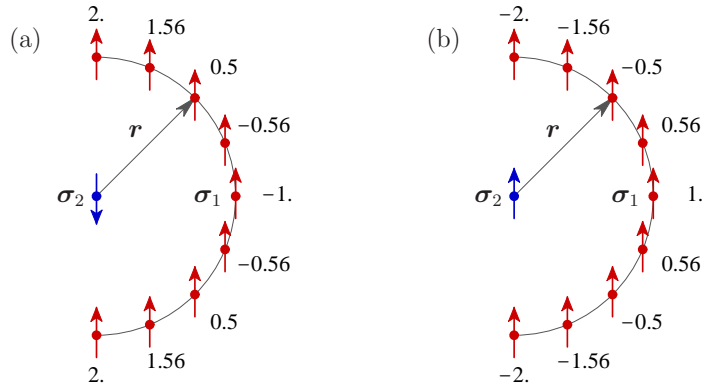}
  \end{center} 
  \caption{Dependence of the tensor interaction energy on the
  spatial orientation of the two spins for fixed antiparallel (a) and 
  parallel spins (b), respectively. The numbers indicate the values of
  $V_{\text{T}}$ for each of the configurations.}
  \label{fig:tensorcorrillu}
\end{figure}

The origin of the tensor correlations is the long-range tensor
interaction, particularly in the $(S,T)=(1,0)$ channel. As seen in Fig.
\ref{fig:deuterondens}, the tensor correlations are revealed through the
dependence of the two-body density on the spatial orientation of the two
nucleons with respect to their spins. This can be understood from the 
structure of the tensor operator  $\SO_{12} = \frac{3}{\rO^2}
(\sigmaOV_1\cdot\rOV)(\sigmaOV_2\cdot\rOV)  -
(\sigmaOV_1\cdot\sigmaOV_2)$. For illustration we consider the interaction
energy $V_{\text{T}} = -[\frac{3}{r^2} (\sigmaV_1\cdot\rV)
(\sigmaV_2\cdot\rV) - (\sigmaV_1\cdot\sigmaV_2)]$, where the minus sign
was introduced to account for the sign of the radial dependence of the
tensor interaction. Spin vectors and relative coordinate enter in the same
way as for the magnetic dipole-dipole interaction. Figure
\ref{fig:tensorcorrillu} gives the values of $V_{\text{T}}$ for different
spatial orientations of two fixed antiparallel and parallel spins,
respectively. For antiparallel spins, those configuration with $\rV$
perpendicular to the spin direction are energetically favoured.
Contrariwise, for parallel spins a fully aligned arrangement of spins and
relative coordinate is preferred. This explains the structure of the
two-body density shown in Fig. \ref{fig:deuterondens}. It is clear that
these tensor correlations between the two spins and the relative
coordinate cannot be described adequately by a single Slater determinant.

The aim of this paper is to devise and apply a method to include central
and tensor correlations explicitly into simple model spaces. We discuss
this approach, the Unitary Correlation Operator Method (UCOM), in Sec.
\ref{sec:ucom}. The structure of the correlated realistic interaction is
presented in Sec. \ref{sec:corrint} with application to the AV18
potential. Section \ref{sec:groundstate} introduces a simple yet powerful
variational model for the treatment of the nuclear many-body problem based
on the Slater determinant of Gaussian wave packets used in Fermionic
Molecular Dynamics (FMD). Ground state calculations up to mass numbers
$A\lesssim60$ are performed, and binding energies, charge radii, and charge
distributions are compared to experiment. Finally, Sec.
\ref{sec:angularmomproj} discusses the projection onto angular momentum
eigenstates of the intrinsically deformed variational states and compares
the resulting rotational spectra with experiment.

\section{Unitary Correlation Operator Method (UCOM)}
\label{sec:ucom}

\subsection{Concept}

The basic idea of the Unitary Correlation Operator Method (UCOM) is the
following: Introduce the dominant short-range central and tensor
correlations into a simple many-body state by means of a state-independent
unitary transformation. The unitary operator $\CO$ describing this
transformation applied to an uncorrelated many-body state $\ket{\Psi}$
leads to a correlated state $\ket{\corr{\Psi}}$
\footnote{Throughout the paper, hats identify correlated quantities;  
upright symbols indicate operators.},
\eq{ \label{eq:corr_state}
  \ket{\corr{\Psi}} = \CO\; \ket{\Psi} \;.
}
In the simplest case, the uncorrelated state is a Slater determinant. The
correlated state $\ket{\corr{\Psi}}$, however, is no longer a Slater
determinant. Due to the complex short-range correlations introduced by the
correlation operator $\CO$, any expansion of $\ket{\corr{\Psi}}$ in a
basis of Slater determinants will require a huge number of basis states. 

The representation of short-range correlations in terms of a unitary
state-inde\-pendent operator is technically very advantageous. Calculating a
matrix element $\matrixe{\corr{\Psi}}{\OO}{\corr{\Psi}'}$ of an operator
$\OO$ with correlated states is equivalent to calculating the matrix
element $\matrixe{\Psi}{\corr{\OO}}{\Psi'}$ of the correlated operator
\eq{ \label{eq:corr_operator}
  \corr{\OO} = \CCO \OO \,\CO
}
using uncorrelated states. For the treatment of the many-body problem it
is generally more convenient to correlate all operators of interest and to
evaluate expectation values or matrix elements with the uncorrelated
states.

In accord with the two types of correlations discussed above, we decompose
the correlation operator into separate unitary operators $\CO_{\Omega}$
and $\CO_{r}$ for tensor and central correlations, respectively,
\eq{ \label{eq:correlator}
  \CO = \CO_{\Omega} \CO_{r}
  = \exp\!\Big[-\ii \sum_{i<j} \gO_{\Omega,ij} \Big]\;
    \exp\!\Big[-\ii \sum_{i<j} \gO_{r,ij} \Big]\;.
}
Each of the correlation operators can be written as an exponential
involving a Hermitian generator. Since the correlations considered here
are induced by a two-body potential, the generators are also assumed to be
two-body operators. The detailed form of the two-body generators $\gO_r$ and
$\gO_{\Omega}$ reflects the structure of the central and tensor
correlations, as discussed in the Sec. \ref{sec:intro}. We will construct
these generators in the following sections.

\subsection{Central Correlations}
\label{sec:central_corr}

The short-range repulsion in the central part of the NN-interaction
prevents the nucleons in a many-body system from approaching each other
closer than the extent of the repulsive core, i.e., the two-body density
matrix exhibits a correlation hole at small interparticle distances. One
way to incorporate these correlations into a simple many-body state is by
shifting each pair of nucleons apart from each other. The shift has to be
distance-dependent since it should only affect nucleon pairs which are
closer than the core radius. 

Using this picture, we can construct the following ansatz for the Hermitian
generator of the corresponding unitary transformation
\eqref{eq:correlator}: radial shifts are generated by the component of
the relative momentum $\qOV  = \frac{1}{2} [\pOV_1 - \pOV_2]$ along the
distance vector $\rOV = \xOV_1 - \xOV_2$ of two particles:
\eq{
  \qO_r 
  = \frac{1}{2} [ \qOV\cdot\tfrac{\rOV}{\rO} 
    + \tfrac{\rOV}{\rO}\cdot\qOV ] \;.
}
To describe the dependence of the shift on the particle distance we
introduce a function $s(r)$ and define the following Hermitian generator
for the central correlation operator $\CO_{r}$ \cite{FeNe98}:
\eq{ \label{eq:central_generator}
  \gO_r 
  = \frac{1}{2} [ s(\rO) \qO_r + \qO_r s(\rO) ] \;.
}

To illustrate the effect of the correlation operator $\cO_r = \exp[-\ii
\gO_{r}]$ in two-body space
\footnote{
  $\cO$ stands for the correlation operator in two-body space, whereas
  $\CO$ indicates the general correlation operator in many-body space. In
  general, small symbols are used for $k$-body operators in $k$-body space,
  capital symbols for operators in many-body space.
}
, we apply it to a two-body state $\ket{\psi} = \ket{\Phi_{\text{cm}}}
\otimes\ket{\phi}$. The correlation operator does not act
on the centre of mass component $\ket{\Phi_{\text{cm}}}$ by construction.
For the correlated relative part $\ket{\corr{\phi}}$ we obtain in
coordinate representation \cite{FeNe98}:
\eq{ \label{eq:central_wavefunc}
  \braket{\rV}{\corr{\phi}} 
  = \matrixe{\rV}{\cO_r}{\phi}
  = \frac{\Rm(r)}{r}\sqrt{\DRm(r)}\;\; 
    \braket{\Rm(r) \tfrac{\rV}{r}}{\phi} \;.
}
Hence, the application of the correlation operator corresponds to a norm
conserving coordinate transformation $\rV \mapsto \Rm(r) \frac{\rV}{r}$
with respect to the relative coordinate. The function $\Rm(r)$ and its
inverse $\Rp(r)$ are connected to the shift function $s(r)$ by the
integral equation
\eq{
  \int_r^{\Rpm(r)} \frac{\dd\xi}{s(\xi)} = \pm 1 
  \;,\qquad
  \Rpm[\Rmp(r)] = r\;.
}
For slowly varying shift functions $s(r)$, the correlation functions are
approximately given by $\Rpm(r) \approx r \pm s(r)$.  The $\Rpm(r)$ are
determined by an energy minimisation in the two-body system, as discussed
in Sec. \ref{sec:optcorr}. 

Since the NN-interaction depends strongly on the spin $S$ and isospin $T$
of the interacting nucleon pair, we decompose the two-body generators
$\gO_{r}$  into a sum of different generators $\gO_{r}^{ST}$ for each of
the four different $(S,T)$ channels
\eq{
  \gO_{r} = \sum_{S,T} \gO_{r}^{ST} \PiO_{ST} \;,
}
where $\PiO_{ST}$ is a projection operator onto the spin $S$ and isospin
$T$ subspace. The correlation operator in two-body space decomposes into a
sum of independent correlation operator for the different channels
\eq{ \label{eq:centralcorr_spiniso}
  \cO_{r} 
  = \exp\!\Big[-\ii \sum_{S,T}\gO_{r}^{ST} \PiO_{ST} \Big]
  = \sum_{S,T} \exp[-\ii \gO_{r}^{ST}]\; \PiO_{ST}
  = \sum_{S,T} \cO_{r}^{ST} \PiO_{ST} \;.
}
This form is very convenient, since it allows us to treat the $(S,T)$
channels separately. For the sake of a concise formulation, we will  not
write out this spin-isospin dependence in the following, but add it when
needed.

\subsection{Tensor Correlations}
\label{sec:tensor_corr}

The strong tensor part of the NN-interaction induces subtle correlations
between the spin of a nucleon pair and their relative spatial
orientation.  In order to describe these correlations, we need to generate
a spatial shift perpendicular to the radial direction. This is done by the
``orbital momentum'' operator
\eq{
  \qOV_{\Omega} 
  = \qOV - \tfrac{\rOV}{\rO}\;\qO_r
  = \frac{1}{2\rO^2} [\,\lOV\times\rOV - \rOV\times\lOV \,] \;,
}
where $\lOV$ is the relative orbital angular momentum operator. Radial
momentum $\tfrac{\rOV}{\rO}\qO_r$ and the orbital momentum 
$\qOV_{\Omega}$ constitute a special decomposition of the relative
momentum operator $\qOV$ and generate shifts orthogonal to each other. The
complex dependence of the shift on the spin orientation is encapsulated in
the following ansatz for the generator $\gO_{\Omega}$ 
\cite{NeFe03}: 
\eq{ \label{eq:tensorgen}
  \gO_{\Omega}  
  = \frac{3}{2}\vartheta(\rO)
    \big[(\sigmaOV_1\!\cdot\qOV_{\Omega})(\sigmaOV_2\!\cdot\rOV) 
    + (\sigmaOV_1\!\cdot\rOV)(\sigmaOV_2\!\cdot\qOV_{\Omega}) \big] 
  = \vartheta(r)\, \tensorRQO \;,
}
where $\tensorRQO = \frac{3}{2} [(\sigmaOV_1\!\cdot\qOV_{\Omega})
(\sigmaOV_2\!\cdot\rOV) + (\sigmaOV_1\!\cdot\rOV)
(\sigmaOV_2\!\cdot\qOV_{\Omega}) ]$. The two spin operators and the
relative coordinate $\rOV$ enter in a similar manner like in the tensor
operator $\tensorO$, however, one of the coordinate operators is replaced
by the orbital momentum $\qOV_{\Omega}$, which generates the transverse
shift. The size and the distance-dependence of the transverse shift is
given by the function $\vartheta(r)$ --- the counterpart to $s(r)$ for the
central correlator. The isospin dependence of the tensor correlator is
implemented in analogy to the spin-isospin dependence of the central
correlator \eqref{eq:centralcorr_spiniso} through projection operators.

Again we illustrate the effect of the correlation operator $\cO_{\Omega} =
\exp[-\ii \gO_{\Omega}]$ by applying it to a two-body wave
function in coordinate representation. For simplicity we use $LS$-coupled
two-body states $\ket{\phi;(LS)J}$. Starting from a pure $L=0$
uncorrelated wave function with $S=1$, $T=0$, and $J=1$, for example, the
tensor correlator generates a superposition of an $L=0$ and an $L=2$
state 
\eqmulti{ \label{eq:tensorcorr_deuteron}
  \matrixe{\rV}{\cO_{\Omega}}{\phi; (0 1) 1}
  =& \cos(3\sqrt{2}\; \vartheta(r))\;\; \braket{\rV}{\phi; (0 1) 1} \\
  &+ \sin(3\sqrt{2}\; \vartheta(r))\;\; \braket{\rV}{\phi; (2 1) 1} \;.
}
The tensor correlation function $\vartheta(r)$ determines the amplitude and
the radial dependence of the D-wave admixture.

\begin{figure}
  \begin{center}
    \includegraphics[width=1.0\textwidth]{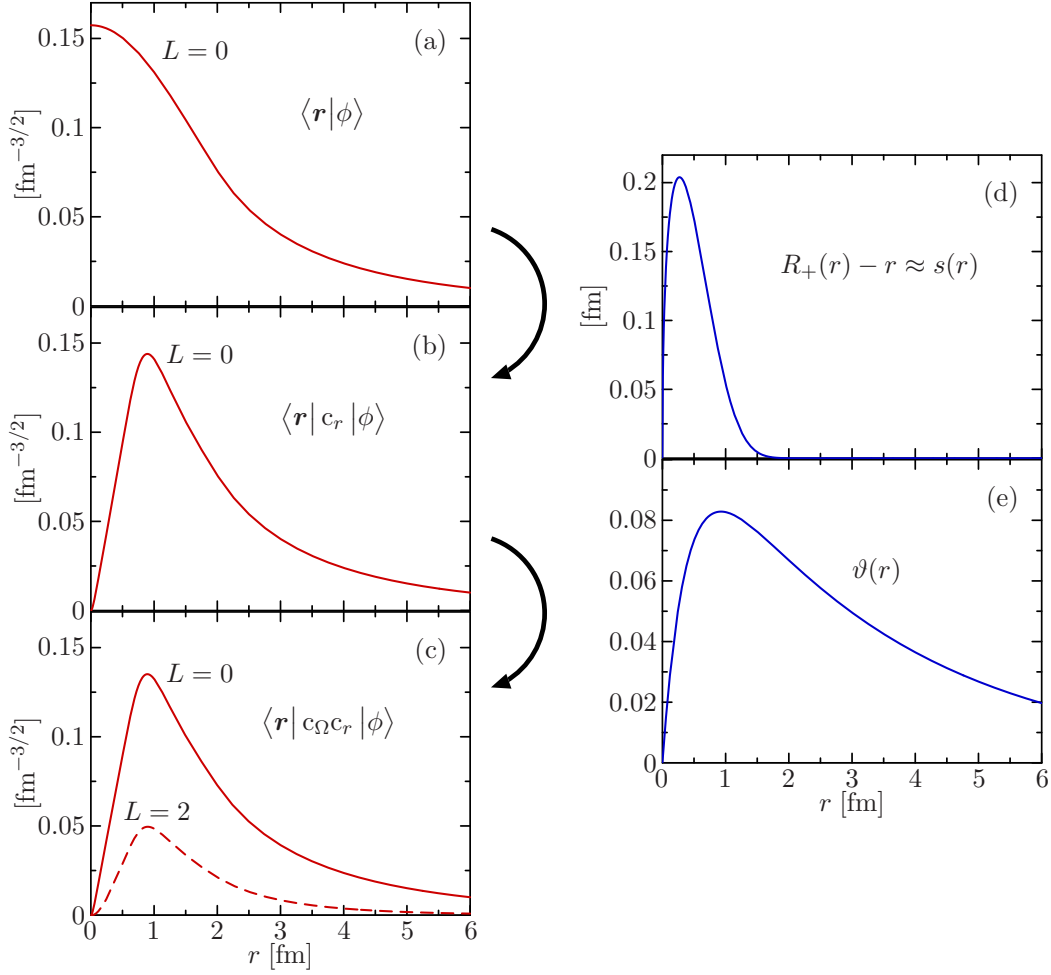}
  \end{center}
  \caption{Construction of the deuteron wave function for the AV18
  potential starting from an uncorrelated wave function shown in panel (a).
  Application of the central correlator with a correlation function
  $\Rp(r)$ shown in (d) leads to the central correlated wave function (b).
  Application of the tensor correlator with $\vartheta(r)$ shown in panel
  (e) produces a D-wave admixture and leads to a realistic deuteron wave
  function depicted in (c).}
  \label{fig:deuteron_corrillu}
\end{figure}

To further illustrate the effect of the central and the tensor correlation
operator on a two-body state, Fig. \ref{fig:deuteron_corrillu} details the
steps from a simplistic ansatz to the exact deuteron wave function for the
AV18 potential. We start with an uncorrelated $L=0$ wave function, shown in
panel (a), which correctly describes the long-range behaviour but does not
contain a correlation hole or a D-wave admixture. Applying, in a first step,
the central correlation operator with the correlation function $\Rp(r)$
depicted in panel (d) leads to the correlated wave function shown in Fig.
\ref{fig:deuteron_corrillu}(b). The amplitude is shifted out of the core
region towards particle distances where the potential is attractive. To
this end, the shift function $s(r)\approx\Rp(r)-r$ has to be large within
the core radius, but has to decrease rapidly outside the core region.  In
a second step, we apply the tensor correlator for a correlation function
$\vartheta(r)$ given in Fig. \ref{fig:deuteron_corrillu}(e). As seen from
Eq. \eqref{eq:tensorcorr_deuteron} this generates a D-wave admixture and
leads to a fully correlated wave function depicted  in panel (c), which is
in nice agreement with the exact deuteron solution. Note that the shape of
the $D$-wave component is determined by the correlation function
$\vartheta(r)$ --- a long-range $L=2$ wave function requires a tensor
correlation function $\vartheta(r)$ of long range.

\subsection{Correlated Operators and Cluster Expansion}
\label{sec:clusterexp}

The explicit formulation of correlated wave functions for the many-body
problem becomes technically increasingly complicated, and the equivalent
notion of correlated operators proves more convenient. The similarity
transformation \eqref{eq:corr_operator} of an operator $\OO$ leads to a
correlated operator which contains irreducible contributions to all
particle numbers. We can formulate a cluster expansion of the correlated
operator
\eq{
  \corr{\OO}
  = \CCO \OO \CO
  = \corr{\OO}^{[1]} + \corr{\OO}^{[2]} + \corr{\OO}^{[3]} + \cdots \;,
}
where $\corr{\OO}^{[n]}$ denotes the irreducible $n$-body part
\cite{FeNe98}. When starting with a $k$-body operator, all irreducible
contributions $\corr{\OO}^{[n]}$ with $n<k$ vanish. Hence, the unitary
transformation of a two-body operator --- the NN-interaction for
example --- yields a correlated operator containing a two-body contribution,
a three-body term, etc. 

The significance of the higher order terms depends on the range of the
central and tensor correlations \cite{FeNe98,NeFe03,Roth00}. If the range
of the correlation functions is small compared to the mean interparticle
distance, then three-body and higher-order terms of the cluster expansion
are negligible. Discarding these higher-order contributions leads to the
two-body approximation 
\eq{
  \corr{\OO}^{C2}
  = \corr{\OO}^{[1]} + \corr{\OO}^{[2]} \;.
}

Already the example discussed in connection with Fig.
\ref{fig:deuteron_corrillu} shows that for the central correlations the
two-body approximation is well justified. The range of the shift function
$s(r)\approx\Rp(r)-r$ is smaller than the mean interparticle distance at
saturation density which is about $1.8\,\text{fm}$. One can show by an
explicit evaluation \cite{Roth00} that higher order contributions due to
central correlations are indeed small for the nuclear many-body problem.

The situation is different for tensor correlations. The tensor correlation
function $\vartheta(r)$ needed to generate the D-wave component of the
deuteron wave function is necessarily long-ranged (see Fig.
\ref{fig:deuteron_corrillu}(e)). In a many-nucleon system, however, the tensor
correlations between two nucleons will not be established up to the same
large distances as in the deuteron. The other nucleons interfere and
inhibit the formation of long-range tensor correlations and thus lead to
an effective screening of the long-range tensor correlations. If we were
to use the long-range tensor correlator suitable for the deuteron also in
the many-body problem, this screening effect would emerge through
substantial higher-order contributions of the cluster expansion.  Since an
explicit calculation of the higher-order tensor contributions to the
cluster expansion is presently not feasible, we will anticipate the
screening of the long-range tensor correlations by restricting the range
of the tensor correlation function \cite{NeFe03}. In this way, we can
improve upon the quality of the computationally simple two-body
approximation. Possible residual long-range tensor correlations that are
not represented in the short-range tensor correlator have to be described
by the many-body states of the model space. We will come back to the
construction of the optimal correlation functions for a given NN-potential
in Sec. \ref{sec:optcorr}.

\section{Correlated Realistic Interactions}
\label{sec:corrint}

\subsection{Nuclear Hamiltonian}

We are now applying the formalism of the Unitary Correlation Operator
Method discussed in Sec. \ref{sec:ucom} to construct the correlated
nuclear Hamiltonian in two-body approximation. The starting point is an
uncorrelated Hamiltonian for the $A$-body system
\eq{ 
  \HO 
  = \TO + \VO
  = \sum_{i=1}^{A} \frac{1}{2 m_N}\, \pOV_i^2
    + \sum_{i>j=1}^{A} \vO_{ij} \;,    
}
consisting of the kinetic energy operator $\TO$ and a two-body potential
$\VO$. For the latter, we employ realistic NN-potentials from the Bonn or 
Argonne family of interactions. Those are given in a closed operator
representation facilitating the use within the UCOM framework. The Argonne
V14 \cite{WiSm84} and the charge independent terms of the Argonne V18
interaction \cite{WiSt95} have the following operator structure
\eqmulti{ \label{eq:argonne}
  \vO_{\text{Argonne}}
  &= \sum_{S,T} \big[v_{ST}^{c}(\rO) + v_{ST}^{l2}(\rO)\;\orbitsqrO \big]\;
  \PiO_{ST} \\
  &+ \sum_{T} \big[v_T^{t}(\rO)\;\tensorO + v_T^{ls}(\rO)\;\spinorbitO
    + v_T^{ls2}(\rO)\;\spinorbitO^2 \big]\; \PiO_{1T} \;,
}
where $\PiO_{ST}$ denotes the projection operator onto spin $S$ and
isospin $T$. In the following, operators $\PiO_S$ with a single index
always refer to a projection in spin-space only. The quadratic spin-orbit
term can be rewritten
\eq{
  \spinorbitO^2 
  = \tfrac{2}{3}\orbitsqrO \PiO_1 - \tfrac{1}{2} \spinorbitO 
    + \tfrac{1}{6}\tensorLLO  \;,
}
where
\eq{
  \tensorLLO 
  = 3 (\sigmaOV_1\cdot\lOV)(\sigmaOV_2\cdot\lOV) 
    - (\sigmaOV_1\cdot\sigmaOV_2)\,\lOV{\,}^2 \;.
}
The non-relativistic configuration space versions of the Bonn potentials 
\cite{Mach89} are parametrised using a different set of
operators 
\eqmulti{\label{eq:bonn}
  \vO_{\text{Bonn}}
  &= \sum_{S,T} \big[v_{ST}^{c}(\rO) + \tfrac{1}{2}\{\qOV^2 v_{ST}^{q2}(\rO) 
    + v_{ST}^{q2}(\rO)\, \qOV^2 \} \big]\; \PiO_{ST} \\
  &+ \sum_{T} \big[v_T^{t}(\rO)\;\tensorO + v_T^{ls}(\rO)\;\spinorbitO
    \big]\; \PiO_{1T} \;.
}
Instead of the $\orbitsqrO$ and $\spinorbitO^2$ terms in \eqref{eq:argonne}, 
the Bonn potentials employ a non-local momentum-dependent term involving
$\qOV^2$. For the following considerations it is convenient to rephrase
this contribution in terms of the radial momentum $\qO_r$ and the angular
momentum $\lOV$
\eq{
 \tfrac{1}{2}\{\qOV^2 v^{q2}(\rO) 
    + v^{q2}(\rO)\, \qOV^2 \}
  = \tfrac{1}{2}\{\qO_r^2 v^{q2}(\rO) 
    + v^{q2}(\rO)\, \qO_r^2 \}
    + \frac{v^{q2}(\rO)}{\rO^2}\; \orbitsqrO \;.
}

\subsection{Tensor Correlated Hamiltonian}
\label{sec:ucom_tensor}

The first step to construct the correlated Hamiltonian in two-body
approximation is the application of the tensor correlation operator 
$\cO_{\Omega} = \exp(-\ii \gO_{\Omega})$ in two-body space. A general way
to evaluate the similarity transformation is by utilising the
Baker-Campbell-Hausdorff expansion
\eqmulti{ \label{eq:BCH_expansion}
  \ccO_{\Omega} \OO\, \cO_{\Omega}
  &= \exp(\ii \gO_{\Omega}) \OO \exp(-\ii \gO_{\Omega}) \\   
  &= \OO + \ii\; \comm{\gO_{\Omega}}{\OO} + \frac{\ii^2}{2}
  \comm{\gO_{\Omega}}{\comm{\gO_{\Omega}}{\OO}} + \cdots \\
  &= \sum_{n=0}^{\infty} \frac{1}{n!} \mathsf{L}_{\Omega}^n\; \OO 
    = \exp(\mathsf{L}_{\Omega})\, \OO \;.
}
In the last line we have introduced a compact notation of the iterated 
commutators in terms of powers of the super-operator 
$\mathsf{L}_{\Omega}\,\circ = \ii \comm{\gO_{\Omega}}{\circ}$ with the
generator $\gO_{\Omega}$ given by Eq. \eqref{eq:tensorgen}. Summing up
the full expansion formally leads to the exponential of the super-operator.

First we study the transformation of the various terms of the realistic
NN-potentials \eqref{eq:argonne} and \eqref{eq:bonn}. A minimal set of
operators  we have to consider in order to formulate the tensor correlated
interaction is $\{\rO, \qO_r^2, \orbitsqrO, \spinorbitO, \tensorO,
\tensorLLO \}$. The distance operator $\rO$ commutes with the generator
$\gO_{\Omega}$, i.e., it is invariant under tensor correlations
\eq{
  \ccO_{\Omega} \rO\, \cO_{\Omega}
  = \rO \;.
}
For the radial momentum $\qO_r^2$, the Baker-Campbell-Hausdorff expansion
terminates after the second order and we obtain a closed expression for
the tensor correlated operator \cite{NeFe03},
\eq{
  \ccO_{\Omega} \qO_r^2 \cO_{\Omega}
  = \qO_r^2 - [\vartheta'(\rO)\,\qO_r 
    + \qO_r \vartheta'(\rO)]\; \tensorRQO
    + [\vartheta'(\rO)\; \tensorRQO]^2 \;,
}
which can be further simplified using the identity $\tensorRQO^2 = 9
[\sOV^2 + 3 \spinorbitO + \spinorbitO^2]$. All other basic operators
require the  evaluation of the full commutator expansion
\eqref{eq:BCH_expansion}. At first order, the following commutators appear
\cite{Neff02}:
\eqmulti{
  \comm{\gO_{\Omega}}{\tensorO}
  &= \ii \vartheta(\rO) [-24\, \PiO_1 - 18\, \spinorbitO + 3\, \tensorO] \\
  \comm{\gO_{\Omega}}{\spinorbitO}
  &= \ii \vartheta(\rO) [-\tensorbarQQO] \\
  \comm{\gO_{\Omega}}{\orbitsqrO}
  &= \ii \vartheta(\rO) [2\,\tensorbarQQO] \\
  \comm{\gO_{\Omega}}{\tensorLLO}
  &= \ii \vartheta(\rO) [7\,\tensorbarQQO] \;,
}
where we have used the short-hand notation
\eq{
  \tensorbarQQO
  = 2\rO^2 \tensorQQO + \tensorLLO - \tfrac{1}{2}\,\tensorO \;.
}
In addition to the original set of operators, a new tensor containing two
orbital momentum operators is generated
\eq{
  \tensorQQO 
  = 3 (\sigmaOV_1\cdot\qOV_{\Omega})(\sigmaOV_2\cdot\qOV_{\Omega})
    - (\sigmaOV_1\cdot\sigmaOV_2)\,\qOV_{\Omega}^2 \;.
}
For the calculation of the second order of the expansion
\eqref{eq:BCH_expansion}, the commutator of $\gO_{\Omega}$ and $\tensorbarQQO$
is required:
\eq{
  \comm{\gO_{\Omega}}{\tensorbarQQO}
  = \ii \vartheta(\rO) [(108 + 96 \orbitsqrO) \PiO_1 + (153 + 36\orbitsqrO)
    \spinorbitO + 15 \tensorLLO ] \;.
}
Again a new operator, $\orbitsqrO\spinorbitO$, emerges, whose
commutator with the generator enters into the third order of the
Baker-Campbell-Hausdorff expansion:
\eq{    
  \comm{\gO_{\Omega}}{\orbitsqrO\spinorbitO}
  = \ii \vartheta(\rO) [-3\;\tensorbarQQO - \hermi{\orbitsqrO \tensorbarQQO}] \;.
}
This in turn leads to the new operator
$\hermi{\orbitsqrO\tensorbarQQO}=\frac{1}{2}(\orbitsqrO \tensorbarQQO +
\tensorbarQQO \orbitsqrO)$, whose
commutator appears in the fourth order of the expansion:
\eqmulti{ \label{eq:last_commutator} 
  \comm{\gO_{\Omega}}{\hermi{\orbitsqrO\tensorbarQQO}}
  =& \ii \vartheta(\rO) [324 \PiO_1 + 477 \spinorbitO + 600 \orbitsqrO 
    + 51 \tensorLLO \\
  &+ 477 \orbitsqrO \spinorbitO + 144 \lOV{\,}^4 
    + 27 \orbitsqrO \tensorLLO + 36 \lOV{\,}^4 \spinorbitO ]  \;.
}
Evidently, with increasing order in the Baker-Campbell-Hausdorff expansion,
operators containing successively higher powers of the angular momentum
$\lOV$ are generated. The last three terms of \eqref{eq:last_commutator}
are of fourth order in $\lOV$ already. In the following, we will neglect
the contributions beyond the third order in angular momentum. Thus we
achieve a closure of the operator set contributing to the
Baker-Campbell-Hausdorff expansion: 
\eqmulti{ \label{eq:operatorset}
  \{\rO, \qO_r^2, &\orbitsqrO, \spinorbitO, \tensorO, \tensorLLO, \\
  &\tensorbarQQO, \qO_r \tensorRQO, \hermi{\orbitsqrO\tensorbarQQO}, 
    \orbitsqrO \spinorbitO  \} \;.
}
We can represent the super-operator $\mathsf{L}_{\Omega}$ defined in
\eqref{eq:BCH_expansion} as a matrix acting on the vector
\eqref{eq:operatorset} of operators. The summation of the full
Baker-Campbell-Hausdorff expansion is then reduced to calculating a matrix
exponential for the super-operator. In this way a closed operator
representation of the tensor correlated potential is constructed.

The set \eqref{eq:operatorset} already contains the operators relevant for
the transformation of the kinetic energy. We can decompose the kinetic
energy $\TO$ in two-body space into a relative, $\tO_{\text{rel}}$, and a
centre of mass contribution, $\tO_\text{cm}$. The latter is not affected
by the correlations. The relative contribution is further decomposed into
a radial and an angular term
\eq{ \label{eq:kinetic_energy}
  \TO 
  = \tO_{\text{cm}} + \tO_{\text{rel}}
  = \tO_{\text{cm}} 
    + \frac{1}{m_N} \Big(\qO_r^2 + \frac{\orbitsqrO}{\rO^2} \Big) \;.
}
Thus the transformation of the kinetic energy can be inferred from the
transformation properties of the first three operators in the set
\eqref{eq:operatorset}.

\begin{figure}
  \begin{center}  
  \includegraphics[width=1\textwidth]{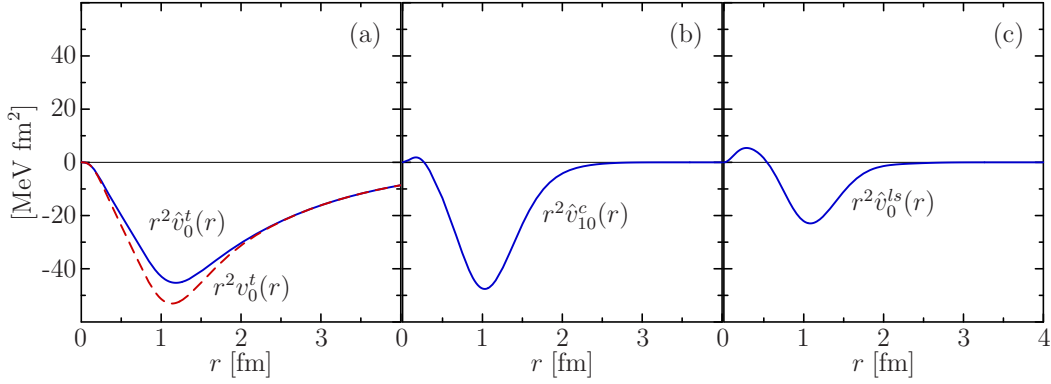}
  \end{center}  
  \caption{Radial dependency of the tensor part of the AV18
  potential  \linemediumdashed[FGRed]\ and the radial dependencies
  \linemediumsolid[FGBlue]\ of the tensor (a), central (b), and
  spin-orbit contribution (c) that result from applying the tensor
  correlator to the isolated tensor part of the interaction. All potentials 
  are for the $(S,T)=(1,0)$ channel, and the appropriate tensor correlation 
  function is given in Sec. \ref{sec:optcorr} ($\alpha$-correlators).} 
  \label{fig:corrtensorint}
\end{figure}

The effect of the tensor correlator on the tensor part of the interaction
is illustrated in Fig. \ref{fig:corrtensorint}. For this example we take
the isolated tensor part of the AV18 potential in the $(S,T)=(1,0)$
channel and apply the tensor correlation operator. 
Through the correlation procedure, i.e., the application of the matrix
exponential according to \eqref{eq:BCH_expansion}, contributions to the
other operator channels \eqref{eq:operatorset} are generated. Most
notably, additional central and spin-orbit contributions emerge, which are
depicted in panels (b) and (c) of Fig. \ref{fig:corrtensorint}. Both, the
induced central and spin-orbit contributions are attractive at
intermediate ranges. Hence, through the correlation procedure, part of the
tensor attraction is transfered to other operator channels, where it can
be exploited using simple uncorrelated many-body states. The marginal
repulsion at short ranges will be completely removed by the central 
correlation operator as discussed in the following.

\subsection{Central and Tensor Correlated Hamiltonian}
\label{sec:ucom_centraltensor}

The subsequent application of the central correlator to the tensor
correlated terms of Hamiltonian is technically simpler. In Section
\ref{sec:central_corr} we have shown that the central correlator acts like
a norm conserving coordinate transformation when applied to a relative
two-body wave function \eqref{eq:central_wavefunc}. Using the
transformation properties of the wave function in coordinate representation
we can immediately derive expressions for the central-correlated operators
in two-body approximation \cite{FeNe98}. The evaluation of the
Baker-Campbell-Hausdorff expansion for the central correlations is
therefore not required.

The application of the central correlation operator $\cO_r$ to the
distance operator $\rO$ confirms the picture of a coordinate
transformation
\eq{
  \ccO_r \rO\, \cO_r 
  = \Rp(\rO) \;,
}
where $\Rp(r)$ is the correlation function introduced in Sec.
\ref{sec:central_corr}. Wherever the distance operator $\rO$
appears in the Hamiltonian, it has to be replaced by a transformed
distance operator $\Rp(\rO)$. This affects, most notably, all radial
dependencies of the different contributions of the interaction
\eq{
  \ccO_r v(\rO)\, \cO_r 
  = v\big(\Rp(\rO)\big) \;.
}
The components of the relative momentum operator $\qOV$ are also influenced
by the central correlations. Using the correlated two-body wave function
\eqref{eq:central_wavefunc}, one can show that
\eq{
  \ccO_r \qO_r \cO_r 
  = \frac{1}{\sqrt{\DRp(\rO)}}\, \qO_r\, \frac{1}{\sqrt{\DRp(\rO)}} 
  \;,\qquad
  \ccO_r \qOV_{\Omega} \cO_r
  = \frac{\rO}{\Rp(\rO)} \qOV_{\Omega} \;.
}
For the quadratic radial momentum $\qO_r^2$, which appears in the tensor
correlated potential as well as in the kinetic energy
\eqref{eq:kinetic_energy}, one obtains
\eq{
  \ccO_r \qO_r^2 \cO_r 
  = \frac{1}{2} \Big\{ \frac{1}{[\DRp(\rO)]^{2}} \qO_r^2 
    + \qO_r^2 \frac{1}{[\DRp(\rO)]^{2}} \Big\}
  + \frac{7 [\DDRp(\rO)]^2}{4 [\DRp(\rO)]^4} - \frac{\DDDRp(\rO)}{2
  [\DRp(\rO)]^3}
  \;.
}
Notice that the transformation of $\qO_r^2$ generates a local potential in
addition to the momentum-dependent term.

All basic operators that act only on the angular part of the two-body
wave function, are invariant under central correlations, e.g.,
\eq{
  \ccO_r \lOV\, \cO_r = \lOV 
  \;,\qquad
  \ccO_r \frac{\rOV}{\rO} \cO_r = \frac{\rOV}{\rO} \;.
}
Utilising the unitarity of the correlation operator, one can easily deduce
from these basic identities the correlated expressions for the composite
tensor and spin orbit operators contained in the operator set
\eqref{eq:operatorset}. One finds that $\orbitsqrO$,  $\spinorbitO$,
$\tensorO$, $\tensorLLO$, $\tensorbarQQO$, $\hermi{\orbitsqrO
\tensorbarQQO}$, and $\orbitsqrO\spinorbitO$ are invariant under similarity
transformation with the central correlation operator.

\subsection{Properties of the Correlated Interaction $\VO_{\textup{UCOM}}$}

Combining the different central and tensor correlated operators discussed
in the previous sections, we can formulate the correlated many-body
Hamiltonian in two-body approximation
\eq{ \label{eq:Hcorr_twobody}
  \corr{\HO}^{C2}
  = \corr{\TO}^{[1]} + \corr{\TO}^{[2]} + \corr{\VO}^{[2]}
  = \TO + \VO_{\UCOM} \;.
}
Its one-body contribution is just the uncorrelated kinetic energy
$\corr{\TO}^{[1]} = \TO$. Two-body contributions arise from the correlated
kinetic energy $\corr{\TO}^{[2]}$ and the correlated potential
$\corr{\VO}^{[2]}$. Together they constitute the correlated interaction
$\VO_{\UCOM}$, which is the basis for the following nuclear
structure studies.

The generic operator structure of the correlated interaction $\VO_{\UCOM}$
is more complicated than that of the bare potentials 
\eqmulti{ \label{eq:VUCOM}
  \vO_{\UCOM}
  &= \sum_{S,T} \big[\corr{v}_{ST}^{c}(\rO) 
    + \{\corr{v}_{ST}^{qr2}(\rO)\, \qO_r^2 \}_{H} 
    + \corr{v}_{ST}^{l2}(\rO)\, \orbitsqrO \big]\; \PiO_{ST} \\
  &+ \sum_{T} \big[ \corr{v}_T^{ls}(\rO)\,\spinorbitO
    + \corr{v}_T^{t}(\rO)\, \tensorO 
    + \corr{v}_T^{tll}(\rO)\, \tensorLLO \\
  &\qquad + \corr{v}_T^{tqq}(\rO)\, \tensorbarQQO 
    + \corr{v}_T^{l2tqq}(\rO)\, \hermi{\orbitsqrO \tensorbarQQO}  \\[6pt]
  &\qquad + \hermi{\corr{v}_T^{qr2trq}(\rO)\, \qO_r^2\, \tensorRQO} 
    + \corr{v}_T^{l2ls}(\rO)\, \orbitsqrO\spinorbitO \big]\; \PiO_{1T} \;,
}
where we use the short-hand notation $\hermi{\cdots}$ to indicate
explicit Hermitisation. The new radial dependencies $\corr{v}_{ST}^{\,\cdots}(r)$
result from the correlation procedure, i.e., from the matrix exponential
for the tensor correlator and the coordinate transformation for the
central correlator. They depend on the radial dependencies of the bare
potential and on the correlation functions. The structure of the different
radial dependencies was discussed in Refs. \cite{FeNe98,NeFe03}. In
summary, the two prime effects of the unitary transformation are:
\emph{(i)} The short-range repulsive core of the central interaction is
removed by the central correlator and an effective repulsion in the
momentum-dependent terms is generated. \emph{(ii)} Additional attractive
central and spin-orbit contributions as well as new tensorial terms are
created by the tensor correlator out of the bare tensor interaction.

Before entering into concrete many-body calculations, we summarise a few
key properties of the correlated interaction $\VO_{\UCOM}$. First, the
correlated interaction is \emph{phase-shift equivalent} to the
uncorrelated potential by construction. This is a direct consequence of
the finite range of the correlation functions. The asymptotics of a
scattering wave function is not altered by the correlation operators and
the phase shifts are preserved. Hence, the unitary correlation operator
provides a tool to generate an infinite manifold of phase-shift equivalent
NN-potentials originating from a single realistic interaction. In addition
to the two-body potential the correlation operator generates a three-body
(and higher-order) interaction which, of course, depends on the particular
correlation functions used. 

Furthermore, there is an interesting connection to the low-momentum
interaction $V_{\text{low}-k}$ determined by means of renormalisation
group techniques \cite{BoKu03}. The momentum space matrix elements of
$\VO_{\UCOM}$ are in very good agreement with the $V_{\text{low}-k}$
matrix elements \cite{NeFe03}. Although both approaches are formally quite
different, the underlying physics is the same: the high-momentum components
of the interaction are treated explicitly --- by the unitary correlation
operator or through the renormalisation group procedure --- leaving an
effective interaction adapted to low-momentum model spaces. One major
practical advantage of the correlated interaction $\VO_{\UCOM}$ is that it
is given in a closed operator representation \eqref{eq:VUCOM}.  Depending
on the particular application, one can easily compute the relevant matrix
elements, e.g., in a plane wave, oscillator, or non-orthogonal Gaussian
basis. The renormalisation group method just provides numerical values for
the momentum-space matrix elements.

\subsection{Optimal Correlation Functions}
\label{sec:optcorr}

A crucial step is the construction of optimal correlation functions
for the use in many-body calculations. These correlation functions depend,
of course, on the bare potential, but they should not depend on the nucleus
under consideration. The aim is to fix a set of optimal
\emph{state-independent} correlation functions which define a fixed
correlated interaction $\VO_{\UCOM}$.  When constructing the correlation
functions one, therefore, has to disentangle state-dependent properties
and state-independent features.

This is an issue in particular for the tensor correlations in the
$(S,T)=(1,0)$ channel. The long-range tensor correlation function
$\vartheta(r)$, which was used in Fig. \ref{fig:deuteron_corrillu}(e) to
reproduce the D-wave admixture of the exact deuteron solution, is certainly
not appropriate for a many-body system. As already discussed in Sec.
\ref{sec:clusterexp}, long-range tensor correlations of a pair of
nucleons are screened through tensor interactions with other nucleons. One
way to isolate the essential state-independent contributions is  to
restrict the tensor correlation functions to short ranges. Long-range
tensor correlations, which are not accounted for explicitly by the
short-range correlator, have to be described through the degrees of
freedom of the many-body states. For the central correlations this
restriction to short ranges emerges automatically, since the strong
repulsive core and the induced correlation hole are short-ranged
themselves. 

Another motivation to consider only correlation functions of 
short range is the validity of the two-body approximation. The two-body
approximation is appropriate as long as the correlation range is small
compared to the mean interparticle distance. In this case, the probability
of finding three nucleons within the range of the correlator is
sufficiently small to neglect three-body and higher-order contributions to
the cluster expansion.

Different methods for the construction of optimal correlation functions
were discussed in detail in Refs. \cite{FeNe98,NeFe03}. Here we employ an
energy minimisation in the two-body system using simple parametrisations
for the central and tensor correlation functions. For the correlation
functions $\Rp(r)$ in the four possible $(S,T)$ channels we will use one of
the following forms:
\eqmulti{
  \Rp^{A}(r)
  &= r + \alpha\, (r/\beta)^{\eta} \exp[-\exp(r/\beta)] \;, \\
  \Rp^{B}(r)
  &= r + \alpha\, (1-\exp[-(r/\gamma)^\eta]) \exp[-\exp(r/\beta)] \;.
}
The tensor correlations functions $\vartheta(r)$ for the two $S=1$
channels are parametrised by
\eqmulti{
  \vartheta^{A}(r)
  &= \alpha\, (1-\exp[-(r/\gamma)^\eta]) \exp[-\exp(r/\beta)] \;, \\
  \vartheta^{B}(r)
  &= \alpha\, (1-\exp[-(r/\gamma)^\eta]) \exp[-r/\beta] \;.
}  
The uncorrelated two-body wave functions are the free zero-energy
solutions of the two-body Schr\"odinger equation with the lowest orbital
angular momentum consistent with antisymmetry for given $(S,T)$. The energy
minimisation is performed for each $(S,T)$ channel separately.
The range of the tensor correlation function in the $(S,T)=(1,0)$ channel
is controlled by using the following integral constraint for the variation 
\eq{ \label{eq:tensor_constraint}
  \int\dd{r}\; r^2 \vartheta(r) 
  = \begin{cases} 
      0.1\, \text{fm}^3 &; \alpha \text{-correlator} \\
      0.5\, \text{fm}^3 &; \gamma \text{-correlator} \\
    \end{cases} \;.
}
Depending on this constraint on the $(S,T)=(1,0)$ tensor correlation
functions, we will refer to the resulting sets of correlators as $\alpha$-
and $\gamma$-correlators, respectively. The parameters of the correlation
functions resulting from the energy minimisation for the AV18 potential
are summarised in Table \ref{tab:optimalcorr}.  The details of their
determination and the corresponding correlation functions for the Bonn A
potential can be found in Ref. \cite{NeFe03}. 

\renewcommand{\arraystretch}{1.2}
\begin{table}
\caption{Parameters of the optimal central and tensor correlation
functions for the AV18 potential as determined in Ref. \cite{NeFe03}.}
\label{tab:optimalcorr}
\begin{center}
\begin{tabular*}{\textwidth}{@{\extracolsep{\fill}} l c c c c c}
\hline\hline
\multicolumn{6}{c}{Central correlation functions $\Rp(r)$} \\
$(S,T)$ & Param. & $\alpha\;\text{[fm]}$ & $\beta\;\text{[fm]}$ 
& $\gamma\;\text{[fm]}$ & $\eta$ \\
\hline
$(0,1)$ & A & 1.379 & 0.8854 &  ---  & 0.3723 \\
$(1,0)$ & A & 1.296 & 0.8488 &  ---  & 0.4187 \\
$(0,0)$ & B & 0.76554 & 1.272 & 0.4243 & 1 \\
$(1,1)$ & B & 0.57947 & 1.3736 & 0.1868 & 1 \\
\hline\hline
\multicolumn{6}{c}{Tensor correlation functions $\vartheta(r)$} \\
$(S,T)$ & Param. & $\alpha$ & $\beta\;\text{[fm]}$ 
& $\gamma\;\text{[fm]}$ & $\eta$ \\
\hline
$(1,0)\alpha$ & A & 530.38 & 1.298 & 1000 & 1 \\
$(1,0)\gamma$ & A & 0.383555 & 2.665 & 0.4879 & 1 \\
$(1,1)$ & B & -0.023686 & 1.685 & 0.8648 & 1 \\
\hline\hline
\end{tabular*}
\end{center}
\end{table}

\begin{figure}
  \begin{center} 
    \includegraphics[width=1\textwidth]{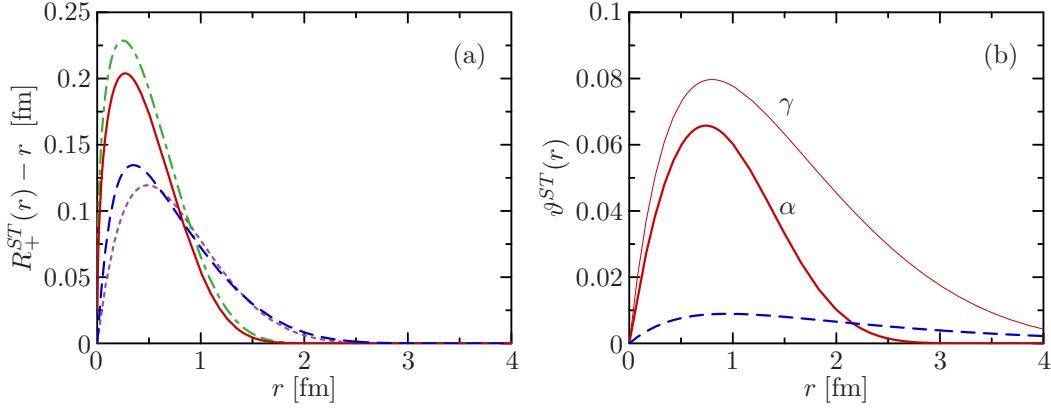}
  \end{center} 
  \caption{Optimal central (a) and tensor correlation functions (b) for the
  AV18 potential. The different lines correspond to the different $(S,T)$
  channels:  $(0,1)$ \linemediumdashdot[FGGreen], 
  $(1,0)$ \linemediumsolid[FGRed], $(0,0)$ \linemediumdotted[FGViolet],
  $(1,1)$ \linemediumdashed[FGBlue].}
  \label{fig:optcorr}
\end{figure}

Plots of the optimal correlation functions for the AV18 potential are
shown in Fig. \ref{fig:optcorr}. All central correlation functions
$\Rp(r)-r$ are of similar short range. Beyond $r\approx1.5\,\text{fm}$ for
the even channels and $r\approx2\,\text{fm}$ for the odd channels the
shift $\Rp(r)-r$ vanishes and the unitary correlation operator acts like
the identity operator. In general, the maximum shift is smaller for the odd
channels, since the uncorrelated $L=1$ wave functions are already depleted
at small $r$ by the centrifugal barrier. The tensor correlation function
$\vartheta(r)$ in the $(S,T)=(1,1)$ channel is very weak and does not lead
to significant effects. For the dominant $(S,T)=(1,0)$ channel, the optimal
tensor correlators for the two different values of the constraint
\eqref{eq:tensor_constraint} are depicted. Only the $\alpha$-correlator
can be considered short-ranged. The $\gamma$-correlator has sizable
contributions beyond $3\,\text{fm}$ and one has to expect significant
three-body contributions to the cluster expansion.

An indirect estimate for higher-order contributions to the cluster
expansion can be obtained by comparing the exact solution of a few-body
problem using the correlated Hamiltonian in two-body approximation with
the solution based on the bare interaction. This has been done in the
no-core shell model \cite{NaVa00b} for \elem{He}{4}
\cite{NeFe03b}. It turns out that for the $\alpha$-correlators the
three- and four-body contributions to the correlated Hamiltonian are
roughly $4\;\text{MeV}$. For the longer ranged $\gamma$-correlators this
contribution grows significantly.

\section{Variational Ground State Calculations}
\label{sec:groundstate}

The correlated realistic interaction $\VO_{\UCOM}$ provides a robust
starting point for different approaches to tackle the many-body problem. 
The dominant short-range correlations are included explicitly though the
unitary correlation operator so that simple low-momentum many-body spaces
--- not able to represent the short-range correlations themselves ---
suffice for a realistic description.

\subsection{Variational Model --- FMD States}
\label{sec:fmd}

Here we will employ a variational model based on an extremely versatile
para\-metrisation for the many-body states which was developed in the
framework of the Fermionic Molecular Dynamics (FMD) model
\cite{Feld90,FeSc97,FeBi95,FeSc00}. The many-body trial state $\ket{Q}$ is
given by a Slater determinant of single-particle states $\ket{q_i}$
\eq{ \label{eq:slaterdet}
  \ket{Q} 
  = \AC (\ket{q_1} \otimes \ket{q_2} \otimes \cdots \otimes \ket{q_A}) \;.
}
The single-particle states $\ket{q}$ are parametrised by Gaussian wave
packets with variable spin orientation and fixed isospin. In general 
a superposition of several wave packets can be used to represent the
single-nucleon state
\eq{
  \ket{q}
  = \sum_{\nu=1}^{n} c_{\nu}\; \ket{a_{\nu}, \bV_{\nu}} \otimes
    \ket{\chi_{\nu}} \otimes \ket{m_{t}} \;.
}
The Gaussian wave packet in configuration space is parametrised in terms of
a complex width parameter $a_{\nu}$ and a complex vector
$\bV_{\nu}=\rhoV_{\nu} + \ii a_{\nu} \piV_{\nu}$, where $\rhoV_{\nu}$ is
the mean position and $\piV_{\nu}$ the mean momentum of the wave packet, 
\eq{
  \braket{\xV}{a_{\nu}, \bV_{\nu}} 
  = \exp\!\Big[-\frac{(\xV-\bV_{\nu})^2}{2 a_{\nu}} \Big] \;.
}
In the simplest case each single particle-state contains $10$ independent
variational parameters. These are determined by a large-scale
numerical minimisation of the expectation value of the correlated
Hamiltonian
\eq{ \label{eq:energyexpect}
  E 
  = \frac{\matrixe{Q}{\corr{\HO}^{C2}-\TO_{\text{cm}}}{Q}}{\braket{Q}{Q}} 
  \to \text{min.}
}
The operator $\TO_{\text{cm}}$ of the centre of mass kinetic energy is
explicitly subtracted to eliminate the centre of mass contribution to the
energy. 

The antisymmetrised Gaussian trial states are extremely flexible and allow
for a multitude of intrinsic structures. Shell-model type states as well
as states with strong $\alpha$-clustering can be described on the same
footing. One should stress that in the following calculations none of
these structures is put in by hand, but that they emerge naturally from
the energy minimisation. A tremendous technical advantage is that all
necessary matrix elements, e.g., for the various terms of the correlated
interaction, can be calculated analytically. This makes large-scale
variational calculations up to mass numbers $A\approx 60$ possible.
Details on the matrix elements and the implementation are given in Ref.
\cite{Neff98}.

\begin{figure}
  \begin{center} 
    \includegraphics[width=0.7\textwidth]{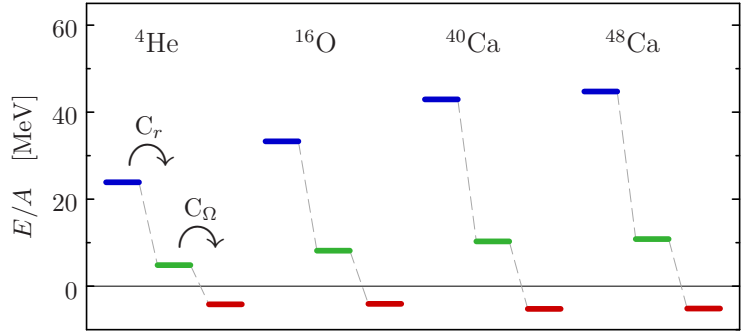}
  \end{center} 
  \caption{Energy expectation value for different nuclei calculated with 
  the uncorrelated Hamiltonian (left), the central-correlated Hamiltonian
  (centre) and the fully correlated Hamiltonian (right) in two-body 
  approximation using the $\alpha$-correlators.}
  \label{fig:energyillu}
\end{figure}

We use this variational model to illustrate the effect of the central
and tensor correlations on the ground-state energy of different nuclei. To
this end we perform the energy minimisation with the fully correlated
interaction \eqref{eq:VUCOM} using the $\alpha$-correlators. With the 
resulting states we then compute the expectation values of the
central-correlated Hamiltonian (without tensor correlations) and of the
uncorrelated Hamiltonian. The results for \elem{He}{4}, \elem{O}{16},
\elem{Ca}{40}, and \elem{Ca}{48} are summarised in Fig. \ref{fig:energyillu}. 
The energy expectation value calculated with the uncorrelated Hamiltonian
is large and positive, i.e., the system is not bound. As discussed
in Sec. \ref{sec:intro} the Slater determinant \eqref{eq:slaterdet} is not
capable of describing short-range central and tensor correlations. This
entails that the repulsive core of the central interaction generates a
large positive energy contribution and that the expectation value of the
tensor part is practically zero.

The proper inclusion of the short-range central correlations lowers the
energy significantly, but still does not lead to bound nuclei since the
attractive contribution of the tensor interaction is missing. This is a
striking demonstration of the importance of the tensor interaction and the
associated correlations. Only after the inclusion of both, central and
tensor correlation operators, we obtain bound ground-states.

We point out that in general all tensor components of the correlated
interaction $\VO_{\UCOM}$ detailed in \eqref{eq:VUCOM} yield negligible
expectation values with a single Slater determinant and can be omitted in
the following. The tensor interaction enters only though the central, the
$\orbitsqrO$, and the $\spinorbitO$ terms generated by the correlation
procedure (cf. Sec. \ref{sec:ucom_tensor}). For the sake of simplicity,
the $\orbitsqrO\spinorbitO$ contribution in \eqref{eq:VUCOM} is replaced
by $2\spinorbitO$ and added to the conventional $\spinorbitO$ term. Thus
two-body states with $L=0$ and $L=1$ are treated correctly and
contributions from $L\geq2$ components, which are suppressed due to the
centrifugal barrier, are approximated.

\subsection{Long-Range Correlations and Three-Body Interactions}
\label{sec:missing_pieces}

\renewcommand{\arraystretch}{1.2}
\begin{table}[b]
\caption{Variational ground state energies and charge radii for the
correlated AV18 potential with the $\alpha$- and $\gamma$-correlators in 
comparison with experiment.}
\label{tab:groundstates}
\begin{center}
\begin{tabular*}{\textwidth}{@{\extracolsep{\fill}} c | c c | c c | c c}
\hline\hline
& \multicolumn{2}{c}{AV18$\alpha$} & \multicolumn{2}{c}{AV18$\gamma$}
  & \multicolumn{2}{c}{Experiment} \\
Nucleus & $E/A\;[\text{MeV}]$ & $r_{\text{ch}}\;[\text{fm}]$
  & $E/A\;[\text{MeV}]$ & $r_{\text{ch}}\;[\text{fm}]$ & $E/A\;[\text{MeV}]$ 
  & $r_{\text{ch}}\;[\text{fm}]$ \\
\hline
\elem{He}{4} & -4.18 & 1.57 & -6.99 & 1.51 & -7.07 & 1.68 \\
\elem{O}{16} & -4.07 & 2.33 & -7.40 & 2.25 & -7.98 & 2.71 \\
\elem{Si}{28} & -3.45 & 2.72 & -6.68 & 2.66 & -8.45 & 3.12 \\
\elem{Ca}{40} & -5.22 & 2.87 & -8.19 & 2.89 & -8.55 & 3.48 \\
\elem{Ca}{48} & -5.14 & 2.87 & -7.87 & 2.93 & -8.67 & 3.47 \\
\hline\hline
\end{tabular*}
\end{center}
\end{table}

In the framework of the FMD variational model we calculate the ground
state energies and the charge radii for a few nuclei using the correlated
AV18 potential with the short-ranged $\alpha$-correlators. Table
\ref{tab:groundstates} summarises the results and compares them to 
experimental data. The magnitude of the binding energies and the charge
radii  for larger nuclei are significantly smaller than the experimental
values. There are two major reasons for these systematic deviations:
long-range correlations and genuine three-body forces.

\emph{Long-range correlations}. Due to the restriction of the range of 
the tensor correlator in the $(S,T)=(1,0)$ channel, long-range tensor
correlations are not adequately described by the correlator. Remember, the
range constraint is used in order to ensure the validity of the two-body
approximation (cf. Sec. \ref{sec:optcorr}). Ideally, the missing
long-range correlations would have to be accounted for by the model-space.
However, in the case of the simple variational model, the single Slater
determinant is not capable of describing long-range tensor correlations.
In principle one could enlarge the model-space such that the  long-range
attraction provided by the tensor interaction can be exploited. This would
lead to an increase in binding energy and radius but is presently not
feasible, at least for larger nuclei. 

Alternatively one could resort to long-range tensor correlators at the
price of including contributions of higher orders of the cluster
expansion. A step into this direction is the use of the longer-ranged
$\gamma$-correlator in the $(S,T)=(1,0)$ channel. The binding energies and
charge radii resulting in two-body approximation are also given in Tab.
\ref{tab:groundstates}. The binding energy increases significantly
compared to the $\alpha$-correlators. Part of this increase is an artifact
of the two-body approximation: The three-body contributions, which are
neglected here, would provide an effective repulsion and compensate most
of the gain in binding energy. Hence, a treatment of long-range tensor
correlations by the Unitary Correlations Operator necessitates a
consistent inclusion of higher-order cluster contributions --- a
technically  extremely challenging task.

\emph{Genuine three-body forces}. As quasi-exact  Green's function Monte
Carlo calculations for small systems show \cite{PiVa02}, realistic
two-body potentials alone do not generate sufficient binding to reproduce
experimental data. This can be remedied by introducing a phenomenological
three-nucleon force which is adjusted to ground states and low-lying
excitations \cite{PiPa01}. Promising developments in effective field theories
\cite{EnMa03} might lead to realistic three-body forces which go
beyond the rather phenomenological three-body terms considered so far in
\emph{ab initio} calculations.

\subsection{Phenomenological Correction}
\label{sec:phencorrection}

The inclusion of three body terms, a genuine three-body force or
three-body contributions of the cluster expansion, poses an enormous
computational challenge. At the moment this is not feasible for the full
range of particle numbers envisioned. We, therefore, resort to a pragmatic
approach and employ a momentum-dependent two-body force to simulate the
missing three-body terms and long-range correlations. The generic
contribution of a momentum-dependent two-body interaction to the energy
per particle in nuclear matter is similar to a local three-body force. 

The structure of the phenomenological correction should be as simple as
possible. For the single Slater determinant employed so far, only the local
and momentum-dependent central terms and the spin-orbit term have sizable
contributions. Therefore, a sensible ansatz for the correction is a
sum of these three terms
\eq{ \label{eq:phencorrection}
  \Delta\vO
  = \Delta v^{c}(\rO) + \qOV\, \Delta v^{qq}(\rO)\, \qOV
  + \Delta v^{ls}(\rO)\, \spinorbitO \;.
}
The symmetric form of the momentum-dependent part is chosen for
convenience and can easily be transformed to the form appearing in
\eqref{eq:VUCOM} or \eqref{eq:bonn}. In order to keep the correction 
simple we consider only spin-isospin independent, i.e., Wigner-type,
forces. Each of the radial dependencies $\Delta v(r)$ is given by a single
Gauss function
\eq{
  \Delta v(r)
  = \gamma \exp[-r^2/(2\kappa)]
}
with strength $\gamma$ and range parameter $\kappa$. The adjustment of the
parameters of the three radial dependencies is done in two steps: First,
the strengths and ranges of the central corrections are adjusted such that
binding energies and charge radii of the doubly magic nuclei \elem{He}{4},
\elem{O}{16}, and \elem{Ca}{40} are in agreement with experiment. Notice
that there are only four parameters available to fit six quantities.
Second, the parameters of the attractive spin-orbit correction are chosen
such that the binding energies of \elem{O}{24} and \elem{Ca}{48} are
consistent with experiment. The resulting correction parameters for the
correlated AV18 potential with the short-ranged $\alpha$-correlators are
\eq{ \label{eq:phencorrection_par}
\begin{array}{r l l}
  \Delta v^{c}(r) :\quad &  
    \gamma=-7.261\;\text{MeV}, & \kappa=2.75\;\text{fm}^2 \;, \\
  \Delta v^{qq}(r) :\quad &  
    \gamma=+14.05\;\text{MeV}\,\text{fm}^2, \quad& \kappa=2.5\;\text{fm}^2 \;,  \\
  \Delta v^{ls}(r) :\quad &  
    \gamma=-2.7\;\text{MeV}, & \kappa=3.0\;\text{fm}^2 \;.
\end{array}
}

It turns out that each of the three terms of the correction plays a
different and physically quite intuitive role. The repulsive
momentum-dependent term is responsible for the reduction of the central
densities and the increase of the charge radii. The attractive local term
provides the bulk of the missing binding energy. The attractive spin-orbit
term gives additional binding especially for nuclei far off stability.
Hence, \eqref{eq:phencorrection} can be considered the simplest possible
form of correction capable of generating all the necessary effects.

One of the effects which is absorbed in the phenomenological correction is
the lack of long-range tensor correlations. From this it is already clear
that the correction depends on the available model space. For an extended
model space, which is capable of describing part of the missing
correlations by itself, the phenomenological correction will be 
different. The parameter set \eqref{eq:phencorrection_par} is adapted to a
model space of states consisting of a single Slater determinant as used in
our variational approach or in Hartree-Fock calculations. If one allows
for superpositions of Slater determinants, e.g., in the framework of
variation after projection or multi-configuration calculations, the
correction has to be readjusted.

\subsection{Binding Energies and Charge Radii}

\begin{figure}
  \includegraphics[width=1\textwidth]{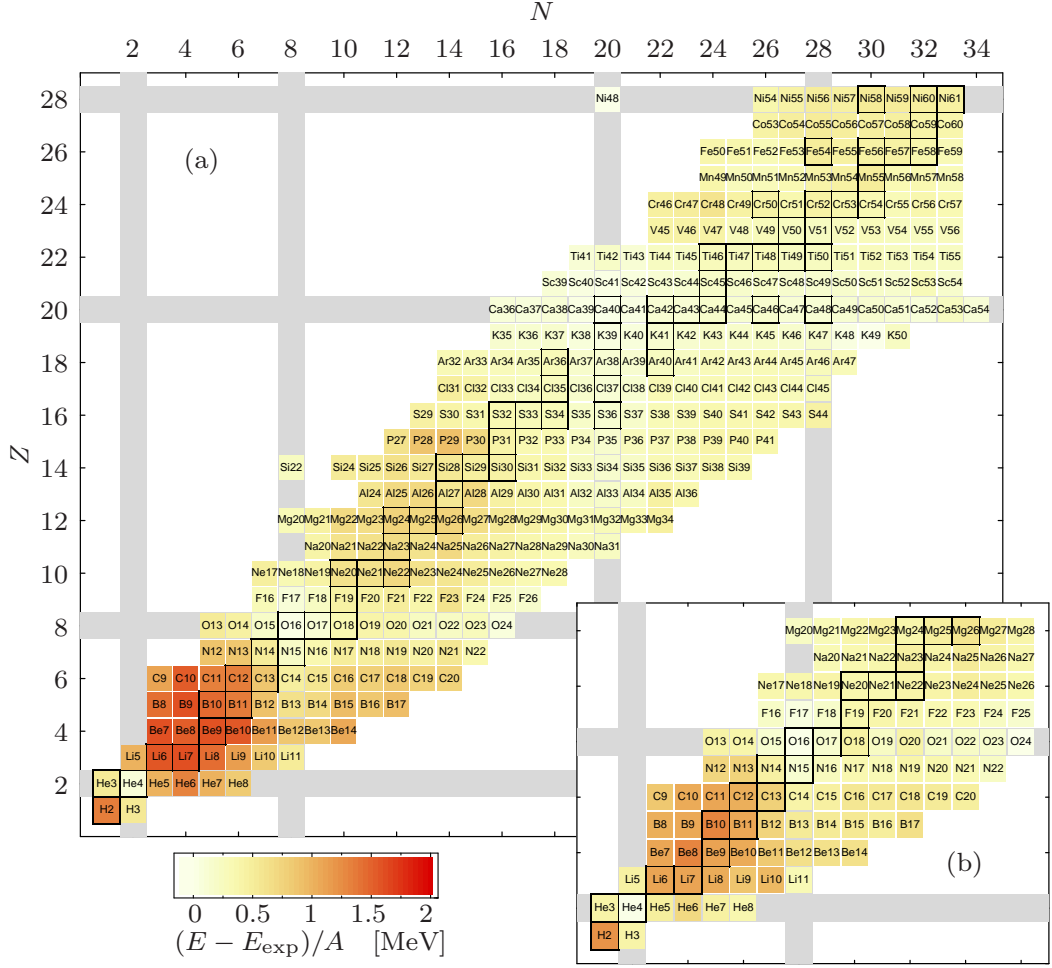}
  \caption{Chart of nuclei accessible to the variational model. The shading
  shows the difference between the variational and the experimental
  binding energy. The main chart (a) was obtained using a single Gaussian wave 
  packet per nucleon. The inset (b) shows results with two Gaussians per nucleon. }
  \label{fig:nuclchart}
\end{figure}

Using the $\alpha$-correlated AV18 potential with the phenomenological
correction \eqref{eq:phencorrection} we perform variational ground state
calculations in the mass region $A\lesssim 60$. Figure \ref{fig:nuclchart}
depicts the part of the nuclear chart accessible to the variational model
based on the FMD trial state \eqref{eq:slaterdet}. The shadings indicate
the deviation of the energy expectation value \eqref{eq:energyexpect} from
the experimental binding energy per nucleon \cite{AuWa95}. The main chart
was obtained using a single Gaussian wave packet to describe the
single-particle states. The agreement with experiment of the absolute
binding energies around the magic numbers and for larger nuclei is quite
good (bright shades). The largest deviations appear for $p$- and
$sd$-shell nuclei away from the shell closures. One reason for these
deviations is the insufficient flexibility of the single-particle trial
states to simulate, e.g., the long-range behaviour of the states
especially for light nuclei. Enhancing the flexibility by using a
superposition of two Gaussians to parametrise the single-particle states
leads to a significant improvement, as the small chart in Fig.
\ref{fig:nuclchart} demonstrates.

\begin{figure}
  \begin{center} 
    \includegraphics[width=1.0\textwidth]{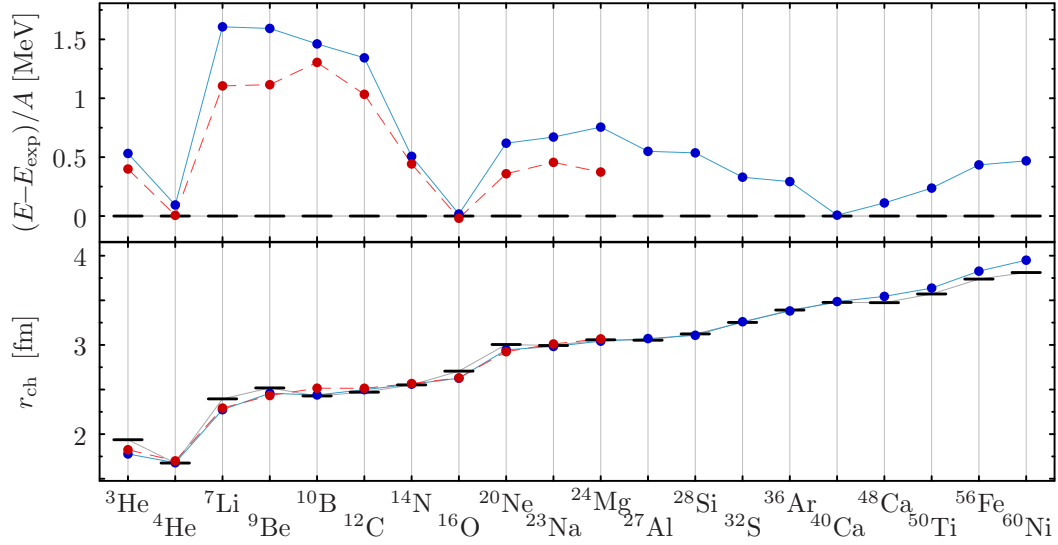}
  \end{center} 
  \caption{Difference between variational and experimental binding energy
  per nucleon (upper panel) and charge radius (lower panel) for selected
  stable isotopes. All calculations are performed using the
  $\alpha$-correlated AV18 potential plus phenomenological correction
  \eqref{eq:phencorrection}. Shown are results obtained with one
  \linemediumsolid[FGBlue]\ and with two Gaussian wave-packets
  \linemediumdashed[FGRed]\ for each single-particle trial state. }
  \label{fig:nucllevels_stab}
\end{figure}

\begin{figure}
  \begin{center} 
    \includegraphics[width=1.0\textwidth]{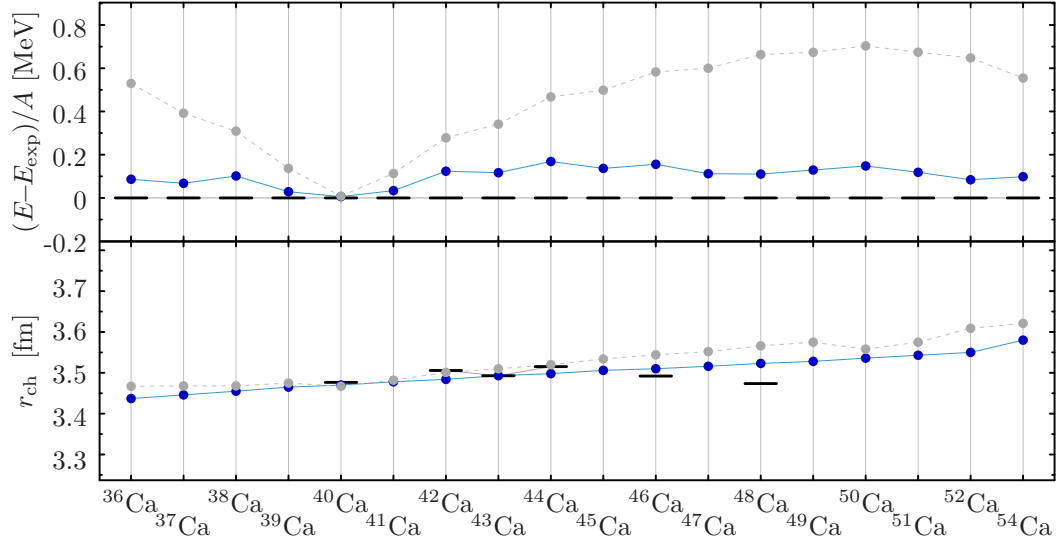}
  \end{center} 
  \caption{Binding energy difference and charge radius for Calcium isotopes
  (cf. Fig. \ref{fig:nucllevels_stab}). Depicted are results based on the
  $\alpha$-correlated AV18 potential with the full phenomenological
  correction \linemediumsolid[FGBlue]\ and with the central terms of the
  correction only \linemediumdotted[FGGray].}
  \label{fig:nucllevels_Ca}
\end{figure}

A more quantitative view is given in Fig. \ref{fig:nucllevels_stab}, where
the binding energy deviations $(E - E_{\text{exp}})/A$ and the charge
radii $r_{\text{ch}}$ are shown for selected stable isotopes. The
calculated charge radii are in excellent agreement with the experimental
data \cite{VrJa87} throughout the whole mass range. The binding energies
are in nice agreement with experiment in the vicinity of the magic nuclei
but start to deviate in between the shell closures, especially for
$p$-shell nuclei. As already mentioned, part of this deviation is due to
an insufficient flexibility of the single-particle trial states. A
refinement of the trial states by using a superposition of two Gaussian
wave-packets (dashed line) instead of one (solid line) leads to a
noticeable  improvement of the energies. The remaining discrepancy can be
reduced further by angular momentum projection as discussed in Sec.
\ref{sec:angularmomproj}. 

For larger nuclei, the trial state with one Gaussian per nucleon  provides
a good description. The binding energy differences and charge radii for
the Calcium isotopes are depicted in Fig. \ref{fig:nucllevels_Ca}. The
binding energies are in very good agreement with experiment, the charge
radii are slightly overestimated for the heavier isotopes but are
generally in agreement with experiment. Figure \ref{fig:nucllevels_Ca}
also illustrates the effect of the spin-orbit contribution in the
phenomenological correction. The gray symbols show results of a
calculation using the $\alpha$-correlated AV18 potential with central
correction but without the spin-orbit term. The spin-orbit term provides
additional attraction for the $N\ne Z$ isotopes and reduces the charge
radii at the same time. The closed-shell $N=Z$ nuclei are not affected.

\subsection{Intrinsic Single-Particle Density Distributions}
\label{sec:density}

Since the variational calculation provides us with the full many-body
wave function we can easily compute other physical quantities, e.g., the
intrinsic single-particle density distribution
\eq{
  \rho(\xV) 
  = \sum_{m_s,m_t} \matrixe{Q}{\adj{\PsiO}_{m_s,m_t}(\xV)
    \PsiO_{m_s,m_t}(\xV)}{Q} \;.
}
Unlike for the two-particle density distribution, the short-range central and
tensor correlations have only marginal influence on the diagonal elements
of the one-particle density matrix \cite{FeNe98,Roth00}. Therefore, we
restrict ourselves to the uncorrelated one-body densities for simplicity.

\begin{figure}
  \begin{center} 
    \includegraphics[width=0.8\textwidth]{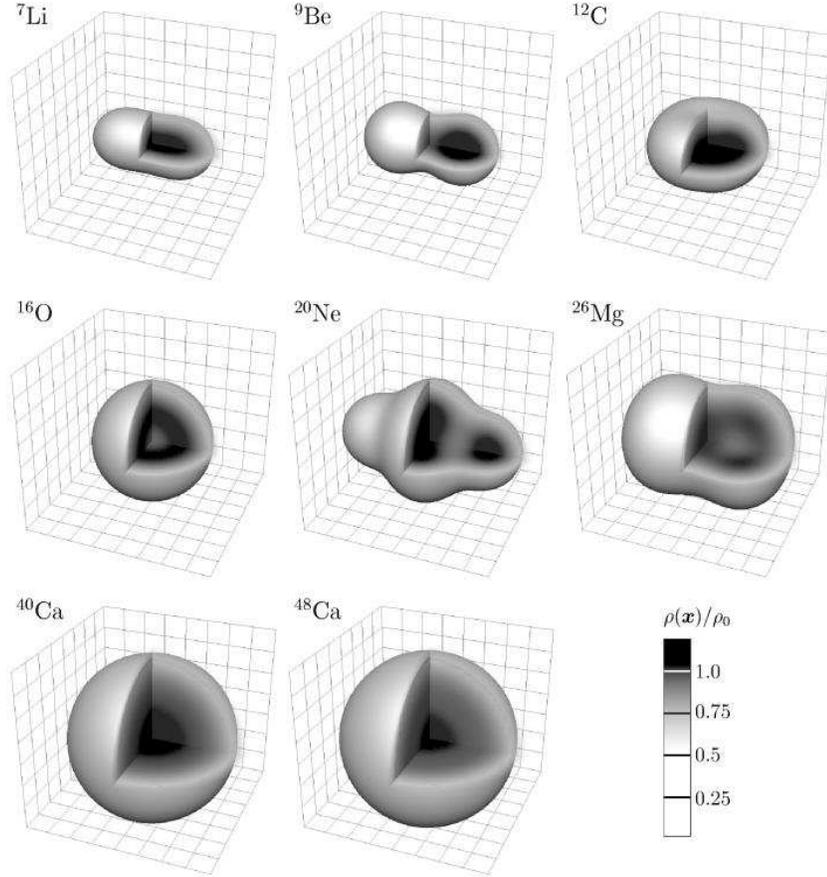}
  \end{center} 
  \caption{Intrinsic total density distributions of several stable
  nuclei. Shown is the iso-density surface for $\rho(\xV)=\rho_0/2$ with one 
  octant cut away to reveal the density distribution in the centre. The
  colour coding on the section gives the density according to the colour
  bar at the right. The background grid has a mesh size of 
  $1\,\text{fm}\times1\,\text{fm}$.}
  \label{fig:density3d_stab}
\end{figure}
\begin{figure}
  \begin{center} 
    \includegraphics[width=1.0\textwidth]{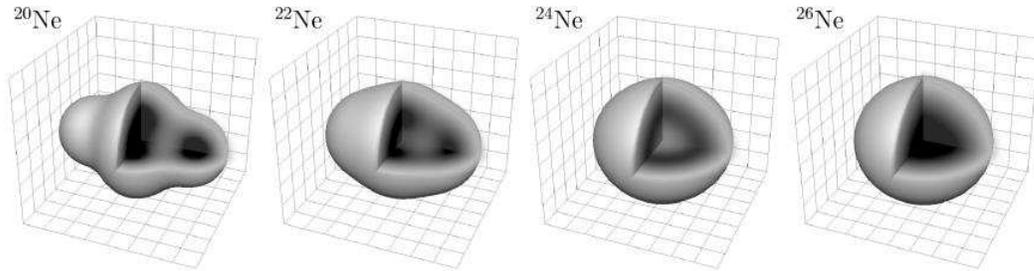}
  \end{center} 
  \caption{Intrinsic single-particle density distribution $\rho(\xV)$ for 
  different Neon isotopes (cf. Fig. \ref{fig:density3d_stab}).}
  \label{fig:density3d_Ne}
\end{figure}

Figure \ref{fig:density3d_stab} shows three-dimensional illustrations of
the intrinsic density distributions for selected nuclei. Depicted is the
iso-density surface for $\rho(\xV)=\rho_0/2$, where
$\rho_0=0.17\;\text{fm}^{-3}$ is the nuclear matter saturation density. In
addition one octant is removed to visualise the interior density
distribution. For the doubly magic nuclei \elem{O}{16}, \elem{Ca}{40}, and
\elem{Ca}{48} the variational calculation leads to spherical symmetric
density distributions in accord with shell-model-type states. For the $p$-
and $sd$-shell nuclei \elem{Li}{7}, \elem{Be}{9}, \elem{C}{12},
\elem{Ne}{20}, and  \elem{Mg}{26} the one-body density distributions
exhibit pronounced  intrinsic deformations. The density distribution of
\elem{Be}{9} reveals a clear two-$\alpha$ substructure; \elem{Ne}{20}
shows a toroidal belt similar to a \elem{C}{12} nucleus supplemented with
two $\alpha$-clusters forming end-caps. For \elem{Mg}{26} only remnants of
$\alpha$-clustering are visible in the rather smooth density distribution
with a multipolar deformation. 

These plots showcase the flexibility of the FMD states. A Slater
determinant of Gaussian single-particle states is capable of describing
shell-model wave functions as well as states with strong
$\alpha$-clustering. We should stress that the structures shown in Fig.
\ref{fig:density3d_stab} are generated by a subtle interplay between the
different terms of the realistic potential. They are not imposed through
constraining the trial state like in $\alpha$-cluster models.

In many cases, strong intrinsic deformations and $\alpha$-clusters present
in a $N\approx Z$ nucleus dissolve gradually when neutrons are added (or
removed). An example are the Neon isotopes depicted in Fig.
\ref{fig:density3d_Ne}. Starting from \elem{Ne}{20} with its
characteristic intrinsic deformation, the addition of neutrons washes out
the $\alpha$-clusters and eventually leads to an almost spherical,
shell-model-type density distribution for $\elem{Ne}{26}$.

\begin{figure}
  \begin{center} 
    \includegraphics[width=0.75\textwidth]{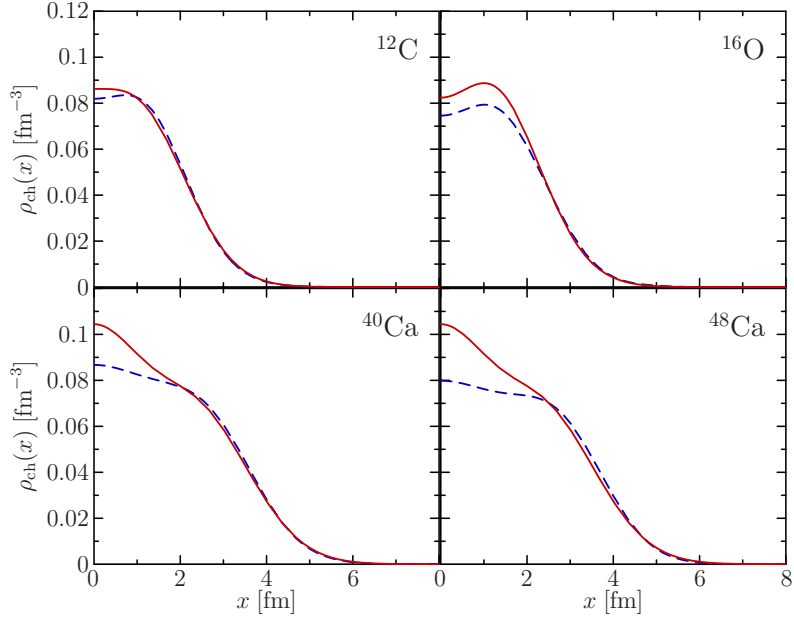}
  \end{center} 
  \caption{Radial charge distribution $\rho_{\text{ch}}(x)$ for different
  nuclei obtained from the variational wave function
  \linemediumsolid[FGRed]\ in comparison with the experimental charge 
  distribution \linemediumdashed[FGBlue]\ resulting from a model-independent 
  analysis \cite{VrJa87}. }
  \label{fig:chargedensity}
\end{figure}

Finally, in Fig. \ref{fig:chargedensity} we compare the radial charge
density distributions for \elem{C}{12}, \elem{O}{16}, \elem{Ca}{40}, and
\elem{Ca}{48} with a model-independent analysis of electron scattering
data \cite{VrJa87}. For this comparison, the exact centre of mass
correction and the proton and neutron form factors are included in the
theoretical curves. The calculated density profiles are in good agreement
with the experimental analysis. The surface structure is reproduced
extremely well. In the interior the modulations of the calculated density
distribution are slightly more pronounced. This behaviour is common to
most models based on a single Slater determinant for the ground state; one
possibility to improve upon this are multi-configuration calculations.
Furthermore, the uncertainties in the experimental determination of the
density profiles are largest in this region.

\section{Angular Momentum Projection}
\label{sec:angularmomproj}

\subsection{Projection after Variation}

Many of the density distributions presented in Sec. \ref{sec:density}
reveal pronounced intrinsic deformations. Evidently the states obtained by
energy minimisation for the trial state \eqref{eq:slaterdet} are not
necessarily angular momentum eigenstates as it is the case for the
eigenstates of the Hamiltonian --- they have to be interpreted as
\emph{intrinsic states}. One can extract angular momentum eigenstates
$\ket{\Psi_{JM}}$ from the intrinsic state $\ket{Q}$ by a standard angular
momentum projection technique \cite{RiSc80}. The angular momentum projected
states are defined through
\eq{ \label{eq:proj_state}
  \ket{\Psi_{JM}} 
  = \sum_K g_K^J\, \PO_{MK}^J\, \ket{Q} \;.
} 
The generalised projection operator $\PO_{MK}^J$ is given by
\eq{
  \PO_{MK}^J 
  = \frac{2J+1}{8\pi^2} \int\dd\omega\;
    D^{J\star}_{MK}(\omegaV)\; \RO(\omegaV) \;,
}
where $D^{J}_{MK}(\omegaV)$ are the Wigner $D$-functions and
$\RO(\omegaV)$ is the unitary rotation operator expressed in terms of
the three Euler angles $\omegaV = (\alpha,\beta,\gamma)$:
\eq{
  \RO(\omegaV)
  = \exp(-\ii \alpha \JO_z) \exp(-\ii \beta \JO_y) \exp(-\ii \gamma \JO_z)
  \;.
}
The coefficients $g_K^{J}$ are formally determined by minimising the
energy expectation value in the projected state $\ket{\Psi_{JM}}$
\eq{ \label{eq:energyexpect_proj}
  E_{J} 
  = \frac{\matrixe{\Psi_{JM}}{\HO-\TO_{\cm}}{\Psi_{JM}}}
    {\braket{\Psi_{JM}}{\Psi_{JM}}}
  = \frac{\sum_{KK'} g_K^{J\star} g_{K'}^{J} h_{KK'}^{J}}
    {\sum_{KK'} g_K^{J\star} g_{K'}^{J} n_{KK'}^{J}} \;,
}
where $h_{KK'}^{J} = \matrixe{Q}{(\HO-\TO_{\cm}) \PO_{KK'}^{J}}{Q}$ and 
$n_{KK'}^{J} = \matrixe{Q}{\PO_{KK'}^{J}}{Q}$. This leads to a generalised
eigenvalue problem for the energies $E_J$ of the projected states and the
coefficients $g_K^{J}$:
\eq{
  \sum_{K'} h_{KK'}^{J}\, g_{K'}^{J}
  = E_{J} \sum_{K'} n_{KK'}^{J}\, g_{K'}^{J} \;.
}

In this paper we employ the angular momentum projection in the framework
of a \emph{projection after variation (PAV)} calculation, i.e., we use the
intrinsic state $\ket{Q}$ obtained by minimising the energy expectation
value \eqref{eq:energyexpect} and subsequently project it onto angular
momentum eigenstates \eqref{eq:proj_state}. 

This scheme is different from a \emph{variation after projection (VAP)}
calculation which is based on a minimisation of the energy expectation
value \eqref{eq:energyexpect_proj} calculated with the projected states. 
The presence of strong intrinsic deformations is associated with a kinetic
energy contribution that makes such states unfavourable when minimising
the intrinsic energy. This energy contribution, however, is removed by the
projection procedure. Therefore, clustering and intrinsic deformation are
generally more pronounced in a VAP framework, where the projected energy
is minimised. In terms of the available model-space, a VAP calculation
goes beyond the variation with a single Slater determinant. The VAP trial
state is a specific superposition of rotated Slater determinants. As
discussed in Sec. \ref{sec:phencorrection} this, in general, necessitates
a readjustment of the phenomenological correction. First results of a
variation after parity projection calculation are available. A full
variation after angular momentum  projection is computationally extremely
involved and will be discussed elsewhere. 

\begin{figure}
  \begin{center} 
    \includegraphics[width=1.0\textwidth]{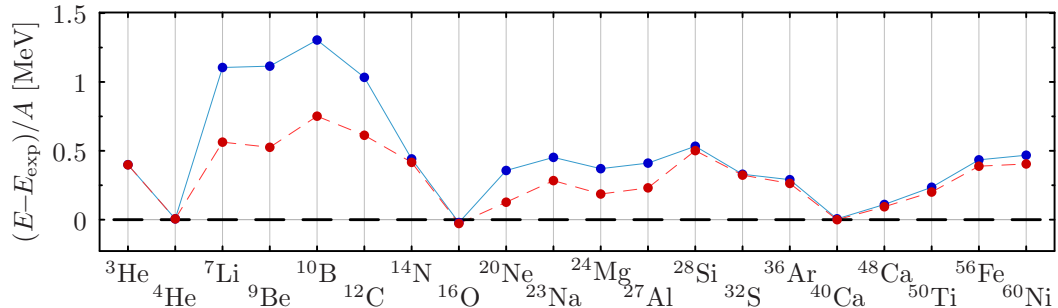}
  \end{center} 
  \caption{Energy deviation $(E-E_{\text{exp}})/A$ for the intrinsic
  variational states \linemediumsolid[FGBlue]\ and for the 
  angular momentum projected intrinsic states (projection after variation) 
  \linemediumdashed[FGRed]. For $A\leq28$ single-particle states 
  composed of two Gaussians are used, heavier isotopes are
  calculated with one Gaussian per nucleon.}
  \label{fig:nucllevels_proj}
\end{figure}

The effect of the angular momentum projection on the ground state energy
is illustrated in Fig. \ref{fig:nucllevels_proj}, where the intrinsic
energies are compared to the energy of the lowest angular momentum
eigenstate obtained in the PAV framework. The deviation from the
experimental binding energies is reduced by a factor $0.5$ for the
intrinsically deformed $p$- and $sd$-shell nuclei. The energy of 
spherical nuclei is not affected. The residual difference has to be
accounted for by further extending the model space, e.g., in the framework
of a multi-configuration calculation.

\subsection{Rotational Spectra}

\begin{figure}
  \begin{center} 
    \includegraphics[width=0.48\textwidth]{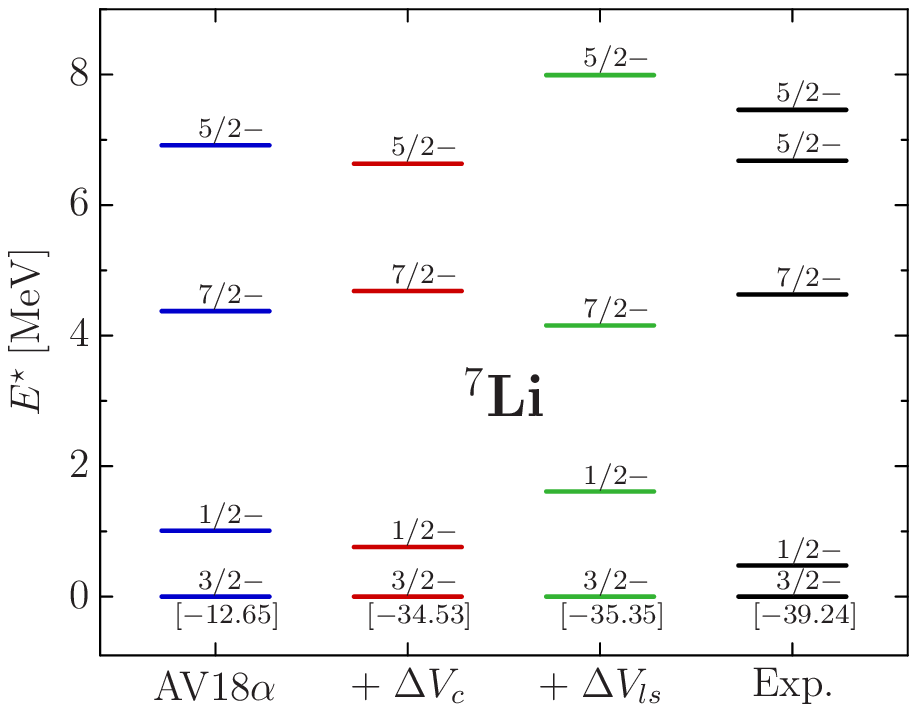} \quad
    \includegraphics[width=0.48\textwidth]{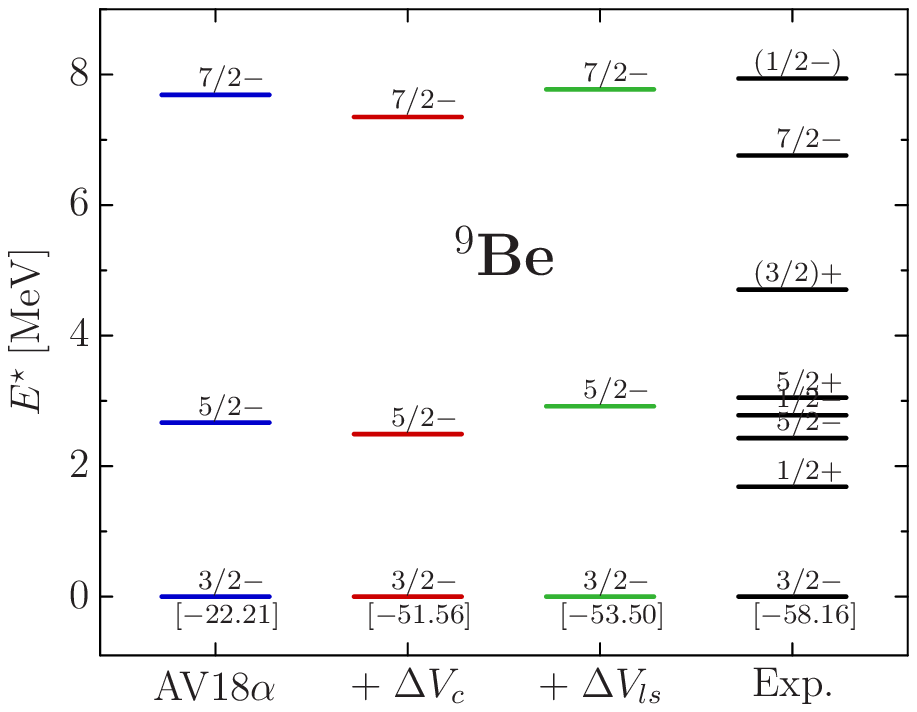} \\[8pt]
    \includegraphics[width=0.48\textwidth]{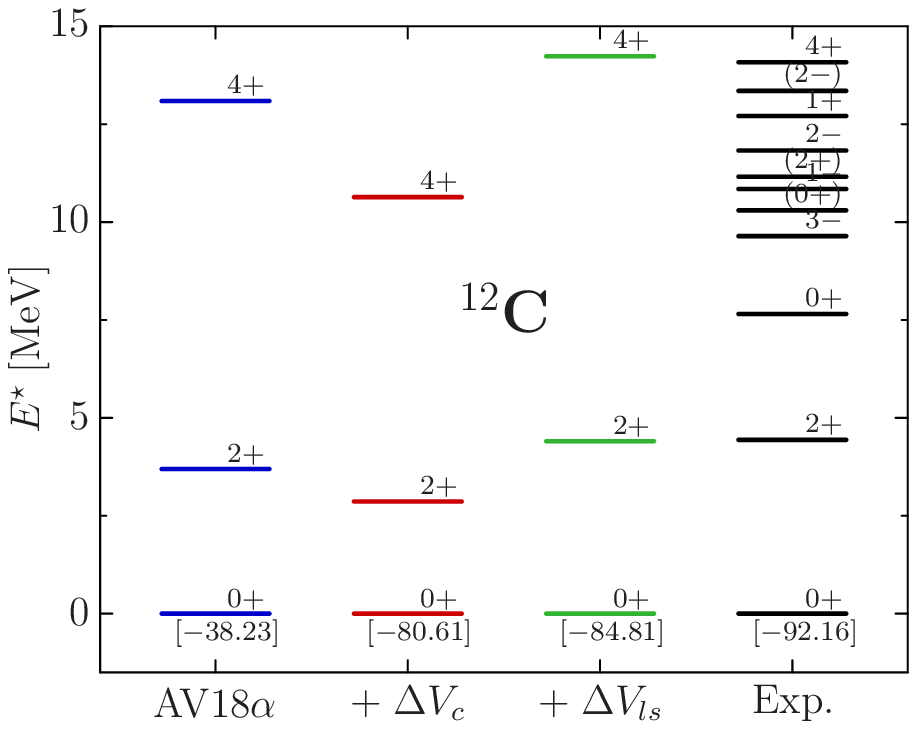} \quad
    \includegraphics[width=0.48\textwidth]{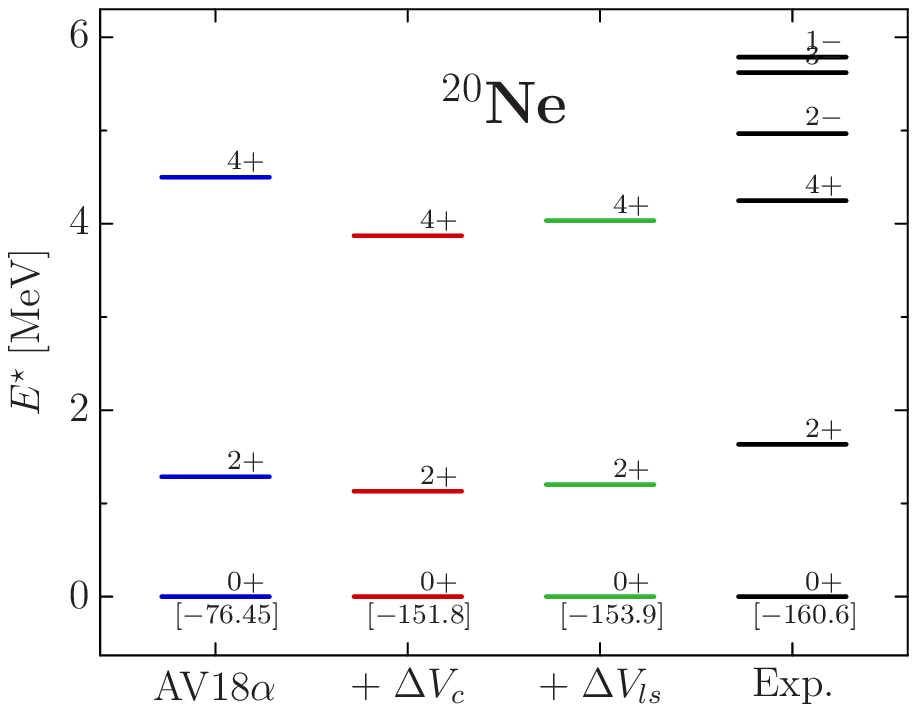} 
  \end{center} 
  \caption{Energy spectra resulting from the angular momentum projection
  of the variational ground states for \elem{Li}{7}, \elem{Be}{9}, 
  \elem{C}{12}, and \elem{Ne}{20}. Shown are the excitation energies
  obtained using the $\alpha$-correlated AV18 potential without correction
  (first column), with central correction only (second column), with
  central and spin-orbit correction (third column) in comparison to the
  experimental spectra (fourth column). The numbers given beneath the
  lowest level are the absolute ground state energies in $[\text{MeV}]$. }
  \label{fig:rotspectra}
\end{figure}

Besides the ground state the PAV procedure provides us with a whole
rotational spectrum based on the intrinsically deformed variational
state.  Figure \ref{fig:rotspectra} shows spectra obtained by angular
momentum projection of the intrinsic states for \elem{Li}{7},
\elem{Be}{9}, \elem{C}{12}, and \elem{Ne}{20}. To give an impression of
the influence of the phenomenological correction to the
$\alpha$-correlated AV18 potential, three different calculations are shown
for each nucleus. First, the $\alpha$-correlated AV18 potential is used
without correction term (left column in each panel). As discussed in Sec.
\ref{sec:missing_pieces} binding energies and charge radii are generally
too small in this case. Nevertheless, the relative spectrum agrees very
well with the experimental one. This is a remarkable result, considering
that for the rotational band the moment of inertia and thus the radius of
the intrinsic density distribution is crucial. 

Second, we perform the variation and projection with the
$\alpha$-correlated AV18 potential including only the central part of the
phenomenological correction \eqref{eq:phencorrection}. Compared to the
significant influence of the correction on the binding energy and the
radius of the ground state the effect on the spectrum is minor (cf. second
column in each of the panels in Fig. \ref{fig:rotspectra}). Generally the 
spectra are slightly compressed due to the increase of the radii induced
by the central part of the phenomenological correction. The inclusion of
the spin-orbit part of the correction stretches the rotational bands again
(see third columns in Fig. \ref{fig:rotspectra}).

Overall, the agreement with experimental data is very encouraging. The
ground state rotational bands of $\elem{C}{12}$ and $\elem{Ne}{20}$ are
reproduced very well. Further improvements are possible in the framework
of multi-configuration calculations, where some of the missing states
corresponding to collective modes can also be described \cite{NeFe04}.

\section{Summary and Outlook}

We have combined two powerful tools for the treatment of the nuclear
many-body problem: the Unitary Correlation Operator Method (UCOM) and the
Fermionic Molecular Dynamics (FMD) model. 

The Unitary Correlation Operator Method (UCOM) provides a systematic way
to derive, from realistic NN-potentials, effective interactions suitable
for simple (low-momentum) model-spaces. Dominant central and tensor
correlations induced by the NN-potential are described explicitly by an
unitary transformation. By applying the unitary correlation operator to a
simple many-body state, e.g., a Slater determinant, the short-range
correlations are imprinted into the state. Alternatively the Hamiltonian
(and all other operators) can be transformed. The resulting  correlated
interaction $\VO_{\UCOM}$ possesses a closed operator representation, is
phase-shift equivalent to the original potential, and leads to
momentum-space matrix elements in accord with the $V_{\text{low}-k}$
matrix elements \cite{NeFe03}. It provides a robust starting point for
many-body methods relying on model spaces not capable of describing
short-range correlations themselves, e.g., variational models with simple
trial states, Hartree-Fock calculations, and multi-configuration
shell-model.

We have employed the correlated interaction $\VO_{\UCOM}$ in the
variational framework of the Fermionic Molecular Dynamics (FMD) model. The
many-body  trial state is a Slater determinant of Gaussian single-particle
states which prove to be extremely flexible. On the same footing, it
allows for the description of spherical shell-model-type wave functions as
well as states with intrinsic deformations and $\alpha$-clustering. At the
same time, the approach is computationally quite efficient so that a
treatment of nuclei up to mass number $A\lesssim60$ is possible for
realistic two-body interactions with complex operator structure. The
inclusion of three-body forces would substantially increase the
computational cost and thus reduce the accessible region of the nuclear
chart. At the present stage, three-body interactions and long-range
correlations are, therefore, simulated by a phenomenological two-body
correction. Charge radii and charge distributions are in nice agreement
with experiment. Around the magic numbers the variational energies agree
well with the experimental binding energies. Away from the shell closures
the intrinsic density distributions exhibit strong deformations and
$\alpha$-clustering, which necessitates projection of the intrinsic states
onto angular momentum eigenstates. Within a projection after variation
(PAV) framework we have achieved very promising results for ground state
energies and rotational spectra.

Natural next steps are variation after projection (VAP) and
multi-configuration calculations. First results along these lines have 
been obtained using a Generator Coordinate Method to implement an
approximate VAP scheme. Constraints on the multipole moments are used to
generate different intrinsic states which serve as input for variation
after projection (w.r.t. the generator coordinates, i.e., the multipole
moments) and multi-configuration calculations. VAP calculations for
\elem{O}{16} show, e.g., that a tetrahedron of $\alpha$-clusters is
energetically more favourable than the spherical shell-model-type state 
resulting from the unprojected variation \cite{NeFe04}. Detailed results
of these investigations will be discussed in a forthcoming publication.

Other future applications of the correlated realistic interaction
$\VO_{\UCOM}$ are Hartree-Fock and RPA calculations. This provides a
consistent scheme to perform nuclear structure calculations based on
realistic NN-potentials also for large nuclei.

\section*{Acknowledgements}

This work is supported by the Deutsche Forschungsgemeinschaft through
contract SFB 634.


\end{document}